\documentclass[journal, letterpaper]{IEEEtran}
\usepackage{cite}

\ifCLASSINFOpdf
\else
\fi

\usepackage{graphicx}
\usepackage{amssymb}
\usepackage{booktabs}
\usepackage{multirow}

\graphicspath{{figures/}}

\usepackage{xcolor}
\definecolor{ECS-Blue}{rgb}{0,0.4392,0.7529}
\definecolor{ECS-Red}{rgb}{0.7529,0,0}
\definecolor{ECS-Green}{rgb}{0.4667,0.5765,0.2353}

\usepackage[para]{threeparttable}
\makeatletter
\def\TPT@doparanotes{\par
   \prevdepth\z@ \TPT@hsize
   \TPTnoteSettings
   \parindent\z@ \pretolerance 8
   \linepenalty 200
   \renewcommand\item[1][]{\relax\ifhmode \begingroup
       \unskip
       \advance\hsize 10em 
       \penalty -45 \hskip\z@\@plus\hsize \penalty-19
       \hskip .15\hsize \penalty 9999 \hskip-.15\hsize
       \hskip .01\hsize\@plus-\hsize\@minus.01\hsize 
       \hskip 0em\@plus .3em
      \endgroup\fi
      \tnote{##1}\,\ignorespaces}%
   \let\TPToverlap\relax
   \def\endtablenotes{\par}%
}

\begin{document}

\setlength{\abovedisplayskip}{5pt plus 2pt minus 0pt}
\setlength{\belowdisplayskip}{5pt plus 2pt minus 0pt}
\setlength{\abovecaptionskip}{0pt plus 2pt minus 0pt}
\setlength{\belowcaptionskip}{0pt plus 2pt minus 0pt}
\setlength{\textfloatsep}{10pt plus 2pt minus 0pt}
\setlength{\floatsep}{5pt plus 1pt minus 0pt}
\setlength{\dbltextfloatsep}{10pt plus 2pt minus 0pt}
\setlength{\dblfloatsep}{5pt plus 1pt minus 0pt}

%
\title{A Family of Current References Based on 2T\\
Voltage References: Demonstration in 0.18-$\mu$m with\\
0.1-nA PTAT and 1.1-$\mu$A CWT 38-ppm/$^\circ$C Designs}
%
%
%
\author{Martin~Lefebvre,~\IEEEmembership{Student Member,~IEEE},
        and David~Bol,~\IEEEmembership{Senior Member,~IEEE}
\vspace{-0.75cm}
\thanks{This work was supported by the Fonds National de la Recherche Scientifique (FNRS), Belgium as part of the European Horizon 2020 project Convergence. \textit{(Corresponding author: Martin Lefebvre.)}
The authors are with the ICTEAM Institute, Université catholique de Louvain (UCLouvain), 1348 Louvain-la-Neuve, Belgium (e-mail: \mbox{martin.lefebvre@uclouvain.be}; david.bol@uclouvain.be).
Color versions of one or more figures in this article are available at
https://doi.org/10.1109/TCSI.2022.3172647.
Digital Object Identifier 10.1109/TCSI.2022.3172647}}

%
%

\markboth{IEEE Transactions on Circuits and Systems--I: Regular Papers, Vol.~xx, No.~x, xx~2022}%
{Shell \MakeLowercase{\textit{et al.}}: Bare Demo of IEEEtran.cls for IEEE Journals}
%

\IEEEoverridecommandlockouts
\IEEEpubid{\begin{minipage}{\textwidth}\ \\[12pt] \begin{scriptsize}This document is the paper as accepted for publication in TCAS-I, the fully edited paper is available at https://ieeexplore.ieee.org/document/9772762. \copyright 2022 IEEE. Personal use of this material is permitted. Permission from IEEE must be obtained for all other uses, in any current or future media, including reprinting/republishing this material for advertising or promotional purposes, creating new collective works, for resale or redistribution to servers or lists, or reuse of any copyrighted component of this work in other works.\end{scriptsize}
\end{minipage}} 


\maketitle

\begin{abstract}
The robustness of current and voltage references to process, voltage and temperature (PVT) variations is paramount to the operation of integrated circuits in real-world conditions. However, while recent voltage references can meet most of these requirements with a handful of transistors, current references remain rather complex, requiring significant design time and silicon area. In this paper, we present a family of simple current references consisting of a two-transistor (2T) ultra-low-power voltage reference, buffered onto a voltage-to-current converter by a single transistor. Two topologies are fabricated in a 0.18-$\mu$m partially-depleted silicon-on-insulator (SOI) technology and measured over 10 dies. First, a 7T nA-range proportional-to-absolute-temperature (PTAT) reference intended for constant-$g_m$ biasing of subthreshold operational amplifiers demonstrates a \mbox{0.096-nA} current with a line sensitivity (LS) of 1.48~$\%$/V, a temperature coefficient (TC) of 0.75~$\%$/$^\circ$C, and a variability $(\sigma/\mu)$ of 1.66~$\%$. Then, two 4T+1R $\mu$A-range constant-with-temperature (CWT) references with (resp. without) TC calibration exhibit a \mbox{1.09-$\mu$A} (resp. \mbox{0.99-$\mu$A}) current with a \mbox{0.21-$\%$/V} (resp. \mbox{0.20-$\%$/V}) LS, a \mbox{38-ppm/$^\circ$C} (resp. \mbox{290-ppm/$^\circ$C}) TC, and a \mbox{0.87-$\%$} (resp. \mbox{0.65-$\%$}) $(\sigma/\mu)$. In addition, portability to common scaled CMOS technologies, such as \mbox{65-nm} bulk and \mbox{28-nm} fully-depleted SOI, is discussed and validated through post-layout simulations.
\end{abstract}
\vspace{-0.1cm}

\vspace{-0.45cm}
%
\IEEEpeerreviewmaketitle

\section{Introduction}
\vspace{-0.05cm}
%
%
%
%
\IEEEPARstart{C}{urrent} and voltage references are crucial components of mixed-signal circuits, ranging from ultra-low-power (ULP) sensor nodes for the Internet of Things (IoT) \cite{Blaauw_2014} to high-performance circuits for compute-in-memory accelerators, such as data converters \cite{Zhang_2017, Yin_2020}. Despite the widely different contexts in which these references are used, they share the common objective of ensuring robustness against PVT variations. On the one hand, recent advances in the design of voltage references have led to simple topologies \cite{Seok_2012, CamposDeOliveira_2018, Fassio_2021}, consisting of only a handful of transistors and consuming a few pW at ambient temperature. While these topologies demonstrate remarkably competitive supply voltage and temperature dependencies with respect to more advanced references, they generally suffer from an increased dependence on process variations and need to be trimmed. Nevertheless, their main advantage lies in the design time reduction arising from the limited number of degrees of freedom. On the other hand, current references remain more complex than their voltage counterparts, often requiring more components and an intricate sizing \cite{Wang_2017}. Nonetheless, recent works have shown that a 2T voltage reference can generate a current once replicated onto a gate-leakage transistor by a 1T buffer \cite{Wang_2018, Zhuang_2020}, suggesting that simpler yet competitive architectures are within reach.\\
\IEEEpubidadjcol
\indent In this paper, we build upon \cite{Wang_2018, Zhuang_2020} to create a family of simple current references without startup circuit sharing two key principles: (i) the generation of a voltage reference by a 2T ULP structure and (ii) its buffering by a single transistor onto a voltage-to-current (V-to-I) converter. Compared with \cite{Wang_2018, Zhuang_2020}, we substitute the gate-leakage transistor by either a self-cascode MOSFET (SCM) or a resistor, respectively resulting in a nA- and $\mu$A-range current. This leads to two novel current reference topologies, fabricated in a \mbox{0.18-$\mu$m} partially-depleted silicon-on-insulator (PDSOI) process. First, a proportional-to-absolute-temperature (PTAT) current reference built by applying a PTAT voltage to an SCM, which is similar to\cite{CamachoGaleano_2005, CamachoGaleano_2008} but with a different PTAT voltage generation. This reference could bias subthreshold operational amplifiers found in ULP sensor nodes with a constant $g_m$. The measured 0.096-nA current has a line sensitivity (LS) of 1.48~$\%$/V, a temperature coefficient (TC) of 0.75~$\%$/$^\circ$C and a $(\sigma/\mu)$ of 1.66~$\%$, while the current reference consumes down to 0.28~nW at 25$^\circ$C. It consists of seven transistors and occupies a silicon area of 8700~$\mu$m$^2$. Second, a constant-with-temperature (CWT) current reference is obtained by the ratio of a voltage and a resistance with matched TCs. It provides a measured \mbox{0.99-$\mu$A} current with an LS of 0.20~$\%$/V, a TC of 290~ppm/$^\circ$C, and a $(\sigma/\mu)$ of 0.65~$\%$, and could be used as a reference for current-steering DACs. It consumes down to 0.64~$\mu$W at 25$^\circ$C, and is made of four transistors and a polysilicon resistor, occupying a total silicon area of 3410~$\mu$m$^2$. Another version with a 4-bit TC calibration reduces temperature dependence down to 38~ppm/$^\circ$C, and provides a \mbox{1.09-$\mu$A} current within a \mbox{4270-$\mu$m$^2$} silicon area.\\
\indent The remainder of this paper is organized as follows. Section~\ref{sec:mos_based_current_generation_techniques} presents existing MOS-based current generation techniques. Then, Section~\ref{sec:a_family_of_current_references} highlights the principles shared by both proposed PTAT and CWT topologies, whose sizing is detailed in Section~\ref{sec:circuit_design_and_sizing_methodology}. Next, simulation and measurement results are discussed in Section~\ref{sec:simulation_and_measurement_results}. We address additional design considerations for an implementation in scaled technologies in Section~\ref{sec:implementation_in_scaled_technologies}. Finally, Section~\ref{sec:comparison_to_the_state_of_the_art} compares our work to the state of the art and Section~\ref{sec:conclusion} offers some concluding remarks.

\section{MOS-Based Current Generation Techniques}
\label{sec:mos_based_current_generation_techniques}
In this section, we describe previous PTAT and CWT current references implemented in a CMOS or BiCMOS technology, based on the four main techniques depicted in Fig.~\ref{fig:1_previous_works}.
\subsection{PTAT Current Reference Topologies}
Among current references, only the self-biased $\beta$-multiplier [Fig.~\ref{fig:1_previous_works}(a)] is generally used for the generation of a PTAT current, achieved by operating transistors $M_{1-2}$ in weak inversion. Indeed, a difference of gate-to-source voltages $\Delta V_{GS}$ proportional to the thermal voltage $U_T$ is applied to the resistor. A common problem is that generating a nA-range current requires a resistance in the order of G$\Omega$, which occupies a significant silicon area. \cite{SerraGraells_2003, CamachoGaleano_2005} have thus proposed to replace the resistor by a 2T SCM, as this structure has a smaller area footprint and generates a current based on the specific sheet current $I_{SQ} \propto T^{2-m}$, where $T$ is the absolute temperature and $m$ is the temperature exponent of the carrier mobility. As $m$ is comprised between 1.2 and 2 in bulk CMOS \cite{Tsividis_1999}, the current can depend quasi-linearly on temperature. Variants of this topology improve the LS through cascoding \cite{CamachoGaleano_2008} or a 1T feedback amplifier \cite{DeLaCruz_2016}. While the PTAT self-biased $\beta$-multiplier only requires a handful of components, the SCM sizing can be tedious because transistors are operated in moderate inversion. However, an SCM can be implemented with transistors from a single threshold voltage ($V_T$) type and reaches a low current within a small area.
\subsection{CWT Current Reference Topologies}
\begin{figure}[!t]
	\centering
	\includegraphics[width=.384\textwidth]{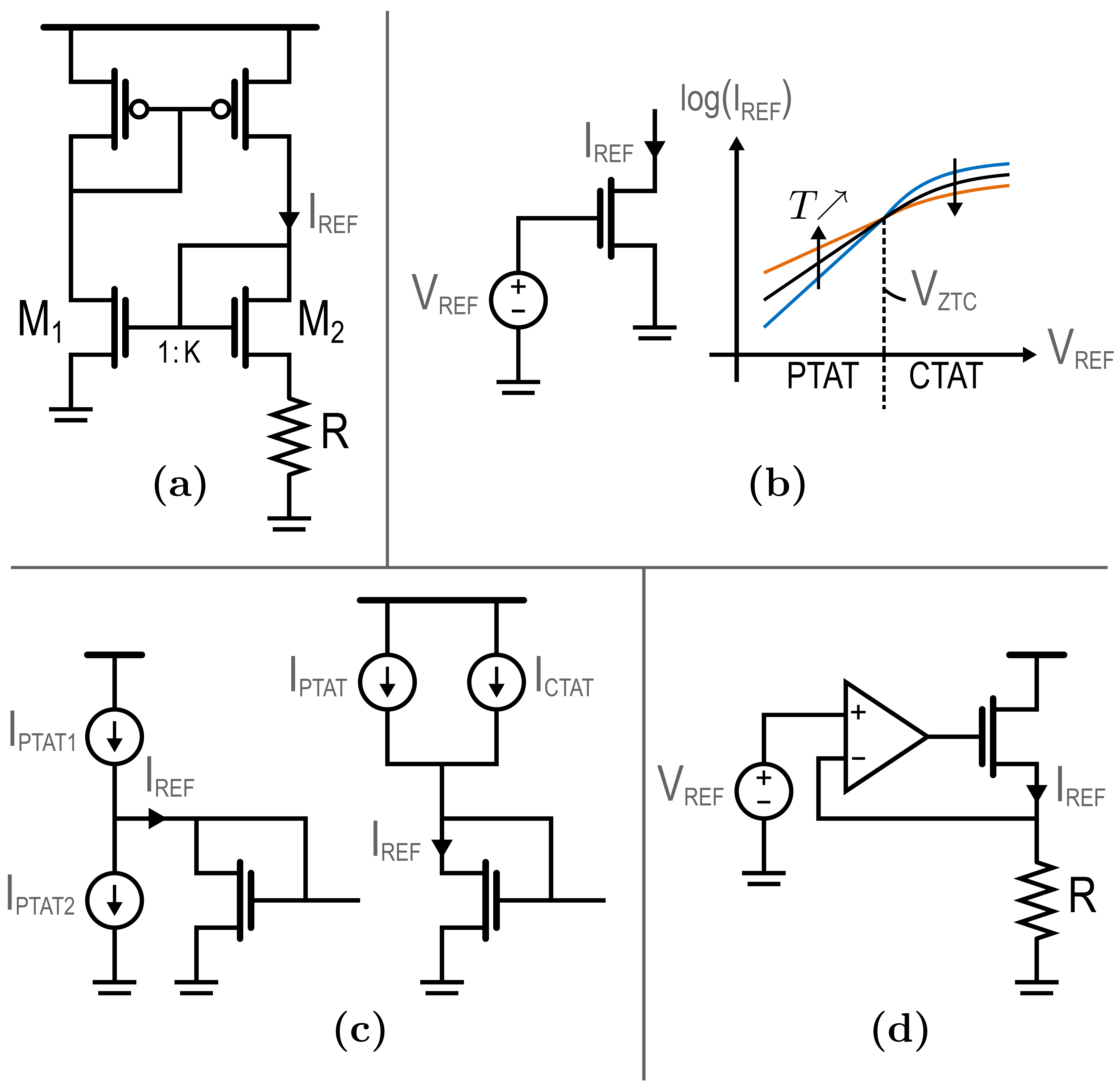}
	\caption{MOS-based current generation techniques can be divided into four categories, mainly oriented towards CWT topologies. (a) Self-biased $\beta$-multiplier, also called $\Delta V_{GS}/R$. (b) Temperature-dependent $V_{GS}$ close to the ZTC, removing the temperature dependence of $I_{DS}$. (c) Subtraction (resp. addition) of two PTAT currents (resp. a PTAT and a CTAT current). (d) Ratio between a reference voltage and a resistance with matched TCs.}
	\label{fig:1_previous_works}
\end{figure}

All four topologies depicted in Fig.~\ref{fig:1_previous_works} can be used to generate a CWT reference current. First, a self-biased $\beta$-multiplier [Fig.~\ref{fig:1_previous_works}(a)] can generate a CWT current with $M_{1-2}$ operating either in weak \cite{Huang_2010, Wang_2019_VLSI, Oguey_1997, Hirose_2005, Osaki_2011} or in strong inversion \cite{Fiori_2005, Kim_2016, Osipov_2019, Azcona_2014, Chouhan_2016}. The strengths of the CWT self-biased $\beta$-multiplier are rather similar to the ones emphasized for the PTAT topologies and are therefore not reminded here.\\
\indent Second, a CWT current can be produced by biasing a transistor close to its zero temperature coefficient (ZTC) with a slightly temperature-dependent $V_{GS}$ [Fig.~\ref{fig:1_previous_works}(b)]. On the one hand, a strong-inversion MOSFET is biased close to its ZTC with either a CWT \cite{Bendali_2007, Ueno_2008} or a slightly PTAT \cite{Dong_2017} gate voltage. On the other hand, a weak-inversion MOSFET requires a slightly complementary-to-absolute-temperature (CTAT) $V_{GS}$ buffered by an OTA \cite{Choi_2014} or a fixed gate voltage and PTAT source voltage, produced by two independent bias circuits \cite{Cordova_2017}. Simpler topologies can operate in weak (resp. strong) inversion with a CTAT (resp. PTAT) gate voltage generated by a 2T ULP voltage reference \cite{Crupi_2018, Fassio_2021_current}. Overall, this current generation technique can be made process-independent with relative ease, by letting the gate voltage track process variations, and has the advantage of a pW-to-nW power consumption. However, a competitive TC can only be achieved close to the ZTC, often resulting in a large reference current. Moreover, complex voltage generators often entail a considerable area overhead.\\
\indent Third, the subtraction of PTAT currents with different TCs [Fig.~\ref{fig:1_previous_works}(c) left], or the weighted sum of PTAT and CTAT currents [Fig.~\ref{fig:1_previous_works}(c) right], can also lead to a CWT current. The subtraction can be done (i) between a PTAT current and its purely temperature-dependent component \cite{Park_2009} or (ii) between two different currents \cite{Far_2015, Yoo_2007, Hirose_2010}. Besides, the weighted sum is performed with a PTAT current, implemented with bipolar transistors as the ratio of a difference of base-to-emitter voltages $\Delta V_{BE}$ and a resistance, and a CTAT current obtained by the ratio of a $\Delta V_{GS}$ \cite{Liu_2010} or a $V_{BE}$ \cite{Yang_2009, Wu_2015} and a resistance. This topology is quite simple, but suffers from several limitations: (i) a calibration of the relative weighting of the currents is necessary to achieve a competitive TC, (ii) the generation of PTAT and CTAT currents can require the use of bipolar transistors, and (iii) two current generators are needed instead of one, resulting in significant area usage.\\
\indent Finally, a CWT current can be obtained as the ratio between a voltage and a resistance with matched TCs [Fig.~\ref{fig:1_previous_works}(d)]. A first category relies on the ratio of a CWT voltage applied either to a CWT resistance \cite{Huang_2020, Wang_2019_TCAS} or to gate-leakage transistors \cite{Wang_2018, Wang_2016}. A second category uses the ratio of a temperature-dependent voltage whose TC matches the one of the resistance. It can be achieved by (i) a slightly PTAT voltage buffered by a linear regulator \cite{Lee_2012}, (ii) a CTAT voltage buffered by a single transistor \cite{Zhuang_2020}, or (iii) a voltage matching the 1$^{\mathrm{st}}$ and 2$^{\mathrm{nd}}$ order TCs of a polysilicon resistor \cite{Wang_2017}. This type of current reference can easily be calibrated to reach a competitive TC performance by matching the 1$^{\mathrm{st}}$ order TC of the resistance, but usually relies on an operational amplifier to buffer the reference voltage onto the resistor, and can lead to rather complex voltage generator architectures if 2$^{\mathrm{nd}}$ order TC matching is required. Obviously, the buffer and voltage generator can result in a considerable power and area overhead.

\section{A Family of Current References}
\label{sec:a_family_of_current_references}
\begin{figure}[!t]
	\centering
	\includegraphics[width=.45\textwidth]{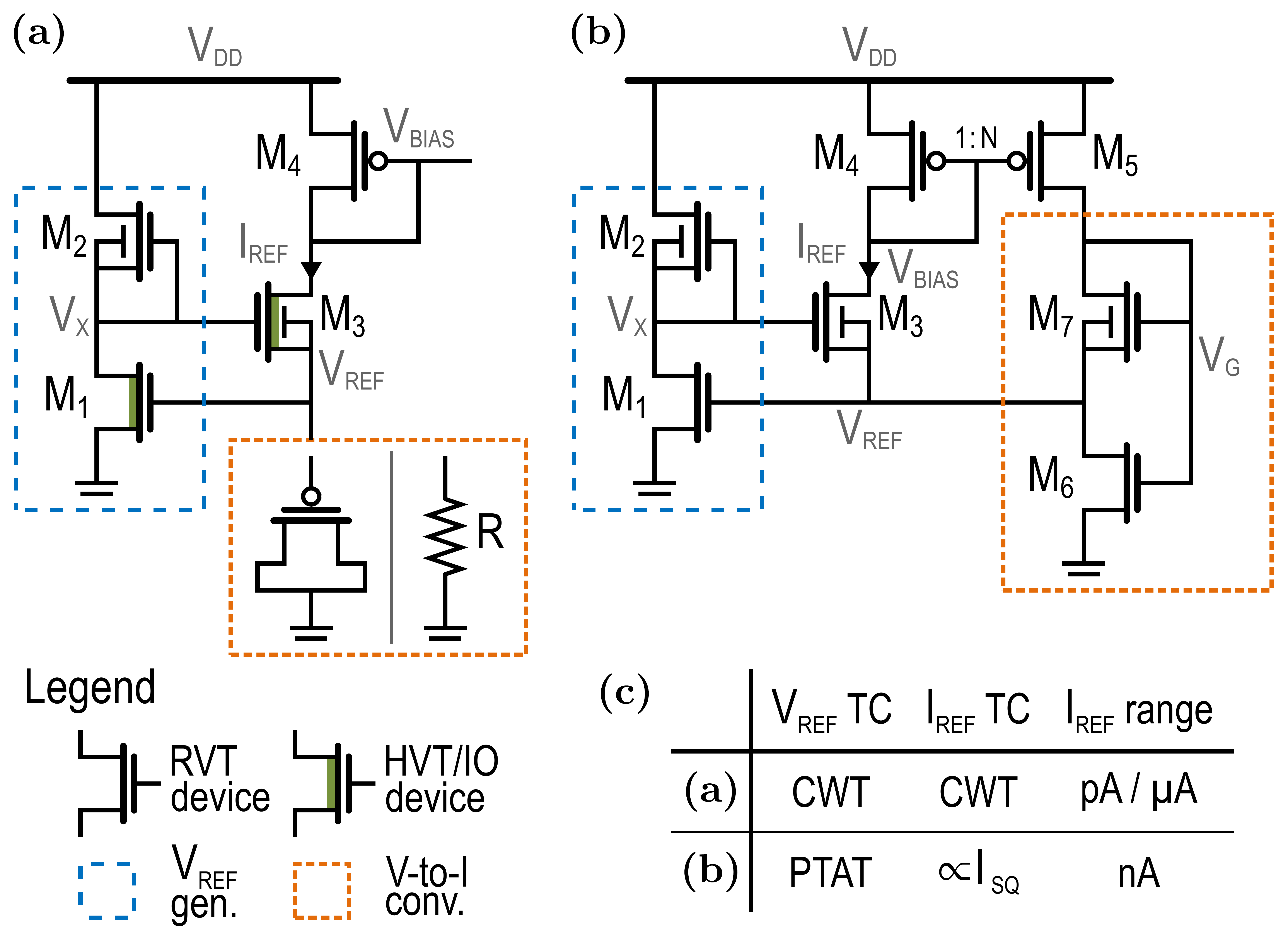}
	\caption{Overview of the family of current references, generating a reference voltage $V_{REF}$ with a 2T ULP structure and converting it into a current by single-transistor buffering onto (a) a gate-leakage transistor \cite{Wang_2018, Zhuang_2020} or a resistor, and (b) an SCM, generating a pA-, $\mu$A- and nA-range current, respectively. Finally, (c) summarizes the TC of $V_{REF}$, as well as the TC and range of $I_{REF}$ across the different architectures.}
	\label{fig:2_family_picture}
\end{figure}
Fig.~\ref{fig:2_family_picture} presents the proposed family of current references, which share the same current generation principle. A reference voltage $V_{REF}$ is generated by the 2T ULP structure formed by $M_{1-2}$, and then buffered onto a V-to-I converter by transistor $M_3$. The range and temperature dependence of the reference current, denoted as $I_{REF}$, are conditioned by the type of transistors used in the voltage reference and the chosen V-to-I converter. First, the current reference in Fig.~\ref{fig:2_family_picture}(a) relies on a CWT voltage reference biasing a gate-leakage transistor or a resistor. The gate-leakage transistor acts as a resistance in the order of G$\Omega$. This structure was proposed in \cite{Wang_2018, Zhuang_2020} and generates a CWT reference current in the pA range. Alternatively, this current reference can rely on a quasi-CWT voltage reference biasing a resistor and compensating for its temperature dependence. It then generates a temperature-independent current in the $\mu$A range. Second, the current reference in Fig.~\ref{fig:2_family_picture}(b) relies on a PTAT voltage reference biasing an SCM and generates a nA-range reference current proportional to the specific sheet current of $M_{6-7}$ \cite{CamachoGaleano_2005, CamachoGaleano_2008}. Fig.~\ref{fig:2_family_picture}(c) summarizes the characteristics of the three references described hereabove, with the $I_{REF}$ range corresponding to the current level that is best suited to each reference, i.e., that leads to a reasonable silicon area. For example, the $\mu$A-range reference could generate a nA-range current, but only at the cost of significant silicon area due to the resistor. In the remainder of this paper, we focus on the nA- and $\mu$A-range current reference topologies, as the pA-range one has already been proposed in prior art \cite{Wang_2018, Zhuang_2020}.\\
\indent The topology brought forward in this work offers three main advantages. First, it requires a small number of transistors, thus reducing the design time by restraining the number of degrees of freedom and potentially resulting in a low area footprint, even though considerations related to local mismatch will limit the area savings. Second, the voltage reference draws a supply current in the pA-to-nA range, resulting in a low power consumption compared to a $\beta$-multiplier or bandgap voltage reference. Third, it does not require any startup circuit as the reference has a unique stable operation point corresponding to a non-zero current, leading to further area savings.

\subsection{2T ULP Voltage Reference}
\label{subsec:a_family_of_current_references_A}
\begin{figure}[!t]
	\centering
	\includegraphics[width=.263\textwidth]{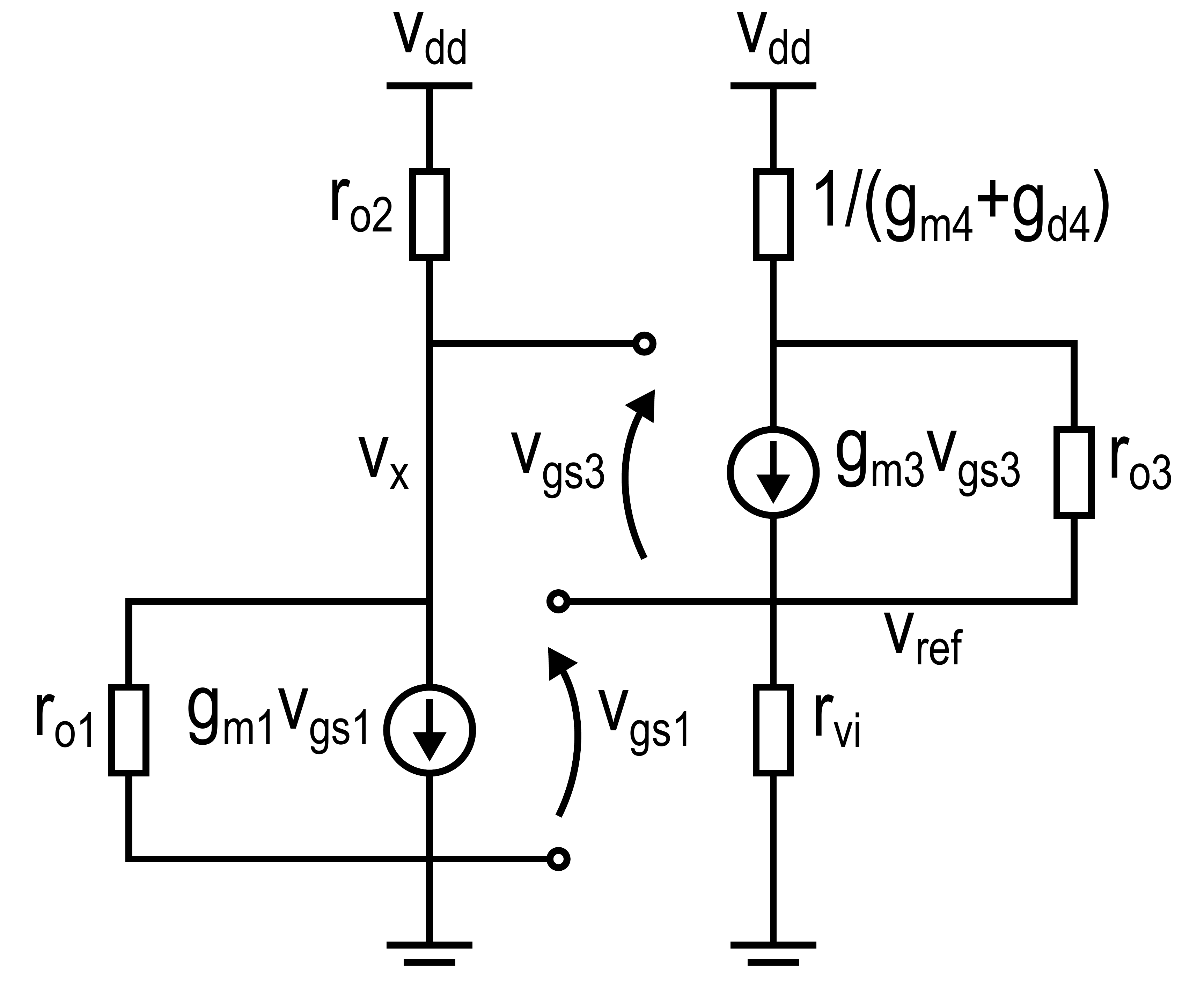}
	\caption{Small signal schematic of the voltage reference, abstracting the V-to-I converter as an equivalent resistance $r_{vi}$.}
	\label{fig:3_small_signal_vref}
\end{figure}
The 2T ULP voltage reference used in this work was proposed in \cite{Wang_2018, Zhuang_2020} in the context of current references and shares the same operation principle as \cite{Adriaensen_2002, Seok_2012, CamposDeOliveira_2018}. It relies on the balance of the subthreshold currents in transistors $M_{1-2}$ to generate a reference voltage. There are three main assumptions underlying the reasoning that follows: (i) for the sake of generality, $M_{1-2}$ have distinct characteristics, (ii) both transistors operate in weak inversion, and (iii) the drain-to-source voltage $V_{DS}$ is larger than $4U_T$, where $U_T$ is the thermal voltage, so the transistors are saturated. Under these assumptions, the drain-to-source current can be expressed as
\begin{equation}
	I_{DS} = I_{SQ}\,S\,\exp\left(\frac{V_{GS}-V_{T0}}{nU_T}\right),\label{eq:ids_wi}
\end{equation}
where $I_{SQ} = \mu C^{'}_{ox} \left(n-1\right) U_T^2$ is the specific sheet current, $\mu$ is the carrier mobility, $C^{'}_{ox}$ is the normalized gate oxide capacitance, $S = W/L$ is the transistor aspect ratio, $V_{T0}$ is the threshold voltage at zero body-to-source voltage $V_{BS}$ and $n$ is the subthreshold slope factor. Applying (\ref{eq:ids_wi}) to $M_{1-2}$ and solving for the reference voltage yields a unique solution
\begin{equation}
	V_{REF} = n_1 U_T \log\left(\frac{I_{SQ2}}{I_{SQ1}}\frac{S_2}{S_1}\right) + \left(\frac{n_2 V_{T01} - n_1 V_{T02}}{n_2}\right).\label{eq:vref}
\end{equation}
This can be explained by the fact that $M_2$ acts as a current source biasing $M_1$ and leads to $V_{REF} > 0$ without requiring a startup circuit. For the $\mu$A-range reference, $M_1$ is chosen to have a larger threshold voltage than $M_2$, i.e., $V_{T01} > V_{T02}$, and the ratio $S_2/S_1$ can be used to tune the reference voltage's TC, whereas for the nA-range current reference, the same transistor type is used for $M_{1-2}$ and (\ref{eq:vref}) simplifies to a purely PTAT voltage
\begin{equation}
	V_{REF} = n U_T \log\left(\frac{S_2}{S_1}\right),\label{eq:vref_PTAT}
\end{equation}
assuming no mismatch between $M_{1-2}$. The LS of the reference voltage is computed from the small signal schematic in Fig.~\ref{fig:3_small_signal_vref}, where $r_{vi}$ denotes the output resistance of the V-to-I converter. First, assuming that $g_{m3} \gg g_{d3}$ and $g_{m4} \gg \frac{1}{r_{vi}}$, the transfer function between $v_{x}$ and $v_{ref}$ is given by (\ref{eq:vx_over_vref}). Then, if $g_{m3} \gg \:\frac{1}{r_{vi}}$, this expression further simplifies to one, which is in line with the common-drain configuration of $M_3$.
\begin{IEEEeqnarray}{RCL}
	\frac{v_x}{v_{ref}} & \simeq & \frac{g_{m3} + g_{d3} + \frac{1}{r_{vi}}}{g_{m3}} \simeq 1\label{eq:vx_over_vref}\\
	\frac{v_{ref}}{v_{dd}} & = & \frac{g_{d2}}{g_{m1} + \frac{v_x}{v_{ref}} \left(g_{d1} + g_{d2}\right)} \simeq \frac{g_{d2}}{g_{m1}}\label{eq:vref_over_vdd}
\end{IEEEeqnarray}
Second, the supply dependence of $V_{REF}$, i.e., $v_{ref}/v_{dd}$, is given by (\ref{eq:vref_over_vdd}) and simplifies to $g_{d2}/g_{m1}$ for $g_m \gg g_d$.

\subsection{nA-Range PTAT Current Reference}
\label{subsec:a_family_of_current_references_B}
To generate a current proportional to the specific sheet current $I_{SQ}$, the nA-range reference relies on a moderate-inversion SCM formed by $M_{6-7}$ and biased by a PTAT voltage, as in \cite{CamachoGaleano_2005}. The use of moderate inversion requires the advanced compact MOSFET (ACM) model \cite{Cunha_1998} to describe the transistor I-V relation. The drain current $I_D$ is given by
\begin{equation}
	I_{D} = I_{SQ} \, S \, \left(i_f - i_r\right),\label{eq:id_ACM}
\end{equation}
where $I_{SQ} = \frac{1}{2}\mu C^{'}_{ox}n U_T^2$ is the ACM specific sheet current, $S$ the transistor aspect ratio, and $i_f$, $i_r$ the forward and reverse inversion levels. The reverse inversion level is non-zero only when the transistor is in triode. Furthermore, the forward inversion level is linked to the gate and source voltages by
\begin{equation}
	V_P - V_S = U_T \left[\sqrt{1 + i_f} - 2 + \log\left(\sqrt{1 + i_f} - 1\right)\right],\label{eq:vp_ACM}
\end{equation}
where $V_P = \frac{V_G - V_{T0}}{n}$ is the pinch-off voltage. A similar relationship is obtained for the reverse inversion level if $V_S$ is replaced by $V_D$ and $i_f$ by $i_r$ in (\ref{eq:vp_ACM}). The ACM equations applied to $M_{6-7}$, respectively in moderate-inversion triode and saturation, give $i_{r6} \simeq i_{f7}$ \cite{CamachoGaleano_2005}. Then, defining $\alpha \triangleq i_{f6}/i_{f7}$, $i_{f7}$ is determined by solving the non-linear equation
\begin{IEEEeqnarray}{CL}
	V_{REF} = & nU_T \Bigg[ \left(\sqrt{1+\alpha i_{f7}}-\sqrt{1+i_{f7}}\right)\IEEEnonumber\\
	& + \log\left(\frac{\sqrt{1+\alpha i_{f7}}-1}{\sqrt{1+i_{f7}}-1}\right) \Bigg].\label{eq:vref_if7}%
\end{IEEEeqnarray}
Finally, the aspect ratios of $M_{6-7}$ must comply with
\begin{equation}
	\frac{S_6}{S_7} \simeq \frac{I_{SQ7}}{I_{SQ6}} \frac{N+1}{N} \frac{1}{\alpha - 1}.\label{eq:S6_over_S7}
\end{equation}
Moreover, the line sensitivity and mismatch of the reference current are directly related to the characteristics of the voltage reference, through the sensitivity $S_{I_{REF}}$. It is computed by means of the chain rule as
\begin{equation}
	S_{I_{REF}} = \frac{1}{I_{REF}} \frac{\partial I_{REF}}{\partial V_{REF}} = \frac{1}{I_{REF}} \frac{\partial I_{REF}}{\partial i_{f7}} \frac{\partial i_{f7}}{\partial V_{REF}},
\label{eq:Siref_vref_PTAT}
\end{equation}
where the variation of $i_{f7}$ with respect to $V_{REF}$ is derived from (\ref{eq:vref_if7}) and $I_{REF} = (I_{SQ7} S_7 i_{f7})/N$ using (\ref{eq:id_ACM}) applied to $M_7$. The sensitivity is thus expressed as
\begin{equation}
S_{I_{REF}} = \frac{2}{i_{f7}nU_T}  \left[\frac{\alpha}{\sqrt{1+\alpha i_{f7}}-1} - \frac{1}{\sqrt{1+i_{f7}}-1}\right]^{-1}.
\label{eq:Siref_PTAT}
\end{equation}
The line sensitivity and mismatch of the reference current can thus be expressed as
\begin{IEEEeqnarray}{RCL}
	\frac{1}{I_{REF}} \frac{\partial I_{REF}}{\partial V_{DD}} & = & S_{I_{REF}} \frac{g_{d2}}{g_{m1}},\label{eq:LS_iref}\\
	\left(\frac{\sigma}{\mu}\right)_{I_{REF}} & = & S_{I_{REF}} \sigma_{V_{REF}},\label{eq:MC_iref}%
\end{IEEEeqnarray}
and depend on both the characteristics of the voltage reference and the sizing of the SCM. Lastly, the minimum supply voltage is given by
\begin{equation}
	V_{DD,\mathrm{min}} = 4U_T + \mathrm{max}(V_{REF} + V_{GS3}, V_{REF} + V_{SG4}, V_G),\label{eq:vddmin_PTAT}
\end{equation}
with each expression in the max function corresponding to one of the branches of the reference depicted in Fig. \ref{fig:2_family_picture}(b).

\subsection{$\mu$A-Range CWT Current Reference}
\label{subsec:a_family_of_current_references_c}
To generate a temperature-independent reference current, the $\mu$A-range current reference relies on a resistor biased by a CWT reference voltage. The reference current is thus given by $I_{REF} = V_{REF}/R$. Assuming that the threshold voltage of transistor $M_i$, denoted as $V_{T0i}$, decreases linearly with temperature $T$ following $V_{T0i}(T) = V_{T0i}(T_0) - \alpha_{V_{T0i}}\left(T-T_0\right)$, and that the resistance temperature variation is captured by $\frac{\partial R}{\partial T}$, the temperature variation of $I_{REF}$ is given by
\begin{IEEEeqnarray}{L}
\frac{\partial I_{REF}}{\partial T} = \frac{1}{R} \left(\frac{nk}{q}\log\left(\frac{I_{SQ2}S_2}{I_{SQ1}S_1}\right) + \left(\alpha_{V_{T01}}-\alpha_{V_{T02}}\right)\right)\IEEEnonumber\\
-\frac{1}{R^2}\left(U_T\log\left(\frac{I_{SQ2}S_2}{I_{SQ1}S_1}\right) + \left(V_{T01} - V_{T02}\right)\right)\frac{\partial R}{\partial T}. \label{eq:diref_dT}%
\end{IEEEeqnarray}
Therefore, the ratio $S_2/S_1$ leading to zero temperature variation of the reference current at $T = T_0$ is
\begin{equation}
\frac{I_{SQ1}}{I_{SQ2}}\exp\left(\frac{q}{nk}\frac{\left(V_{T01}-V_{T02}\right)\frac{1}{R}\frac{\partial R}{\partial T} + \left(\alpha_{V_{T01}}-\alpha_{V_{T02}}\right)}{1 - \frac{T_0}{R}\frac{\partial R}{\partial T}}\right).
\label{eq:S2_over_S1_iref_CWT}
\end{equation}
Similarly to the PTAT current reference, the line sensitivity and mismatch are expressed by (\ref{eq:LS_iref}) and (\ref{eq:MC_iref}), where $S_{I_{REF}}$ is expressed as $1/V_{REF}$. Interestingly, these quantities only depend on the characteristics of the voltage reference. Finally, the minimum supply voltage is given by (\ref{eq:vddmin_PTAT}) excluding $V_G$.

\section{Circuit Design and Sizing Methodology}
\label{sec:circuit_design_and_sizing_methodology}
In this section, we explain the design choices and detail the steps of the sizing methodology used to implement the proposed references in the XFAB \mbox{0.18-$\mu$m} PDSOI technology.

\subsection{nA-Range PTAT Current Reference}
\label{subsec:circuit_design_and_sizing_methodology_A}
The sizing of the PTAT current reference illustrated in Fig.~\ref{fig:2_family_picture}(b) has a twofold objective: minimizing the LS and variability of $I_{REF}$, while achieving the reference current target in nominal conditions (TT process, 25$^\circ$C). Four main parameters can be tuned to achieve this objective:
\begin{enumerate}
	\item $S_2/S_1$, the ratio of $M_{1-2}$ aspect ratios, which amounts to $W_2/W_1$ if these transistors have the same length;
	\item $m$, the multiplier or number of parallel devices used to implement both $M_1$ and $M_2$;
	 \item $\alpha = i_{f6}/i_{f7}$, the ratio of $M_6$ and $M_7$ inversion levels;
	 \item $N$, the multiplicative factor in the current mirror formed by $M_4$ and $M_5$.
\end{enumerate}
\begin{table}[!t]
\centering
\caption{Properties of maximum-length 25-$\mu$m transistors in the XFAB 0.18-$\mu$m PDSOI technology, at 25$^\circ$C. $I_{SQ}$ is determined in the sense of the ACM model.}
\label{table:transistor_properties}
\begin{scriptsize}
\begin{tabular}{lcccc}
	\toprule
	Transistor type & $n$ & $I_{SQ}$ [nA] & $V_{T0}$ [V]\\
	\midrule
	LVT nMOS & 1.21 & 99.63 & 0.433\\
	LVT pMOS & 1.14 & 23.98 & -0.383\\
	\midrule
	RVT nMOS & 1.29 & 67.10 & 0.668\\
	RVT pMOS & 1.41 & 29.26 & -0.749\\
	\bottomrule
\end{tabular}
\end{scriptsize}
\end{table}
\begin{figure}[!t]
	\centering
	\includegraphics[width=.424\textwidth]{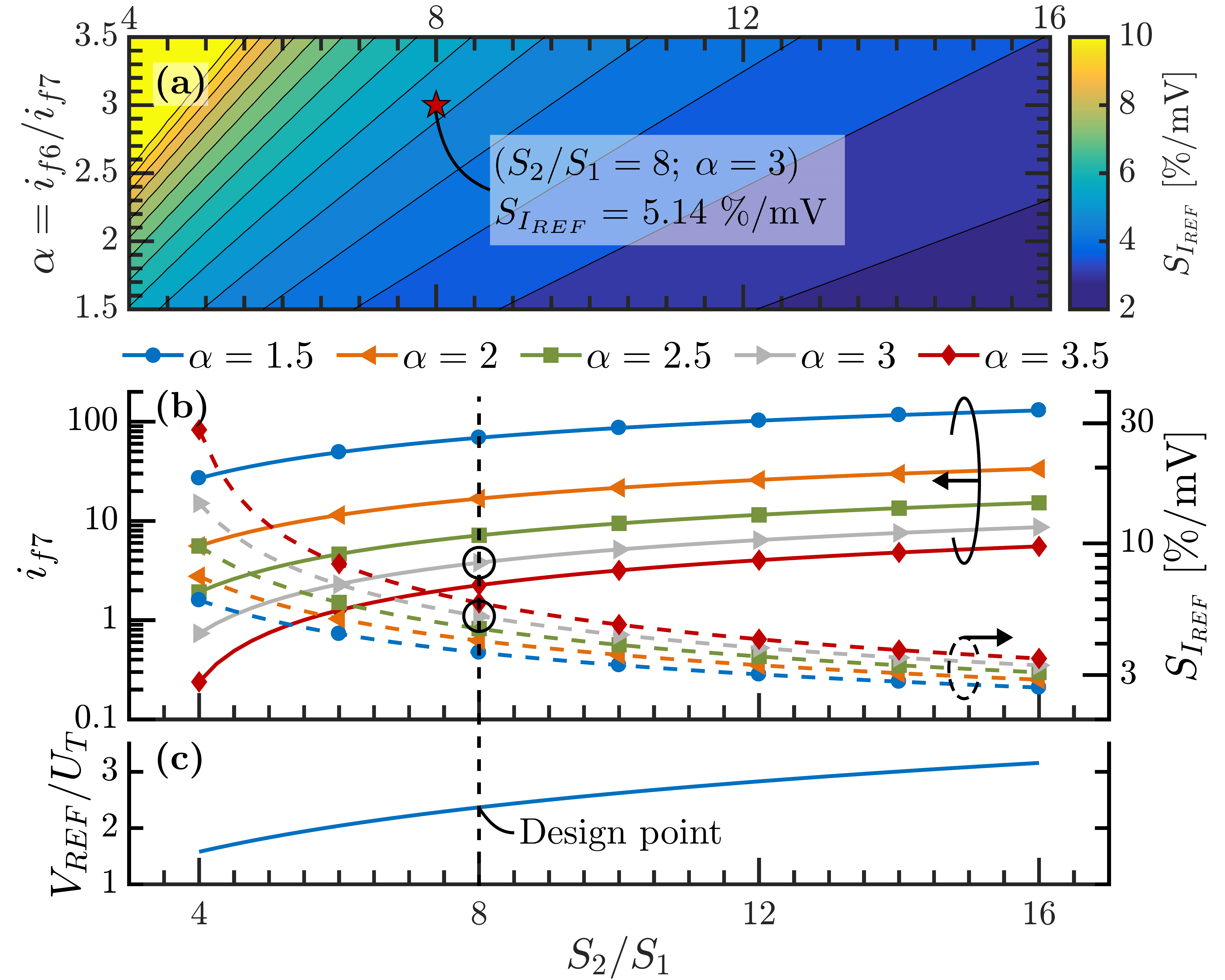}
	\caption{(a) $S_{I_{REF}}$ is dependent on two key design parameters, namely $S_2/S_1$, related to the 2T ULP voltage reference, and $\alpha = i_{f6}/i_{f7}$, related to the SCM. The red star indicates the chosen design point. (b) Inversion level $i_{f7}$ and sensitivity $S_{I_{REF}}$, for different values of $\alpha$ and (c) reference voltage $V_{REF}$, normalized by the thermal voltage $U_T$, as a function of $S_2/S_1$.}
	\label{fig:4_ptat_design}
\end{figure}
The first two degrees of freedom are linked to the 2T voltage reference, while the two latter ones are related to the SCM. We will first explain the sizing of the V-to-I converter, i.e., the SCM, and will then turn to the 2T voltage reference. Besides, low-$V_{T}$ (LVT) devices are selected to implement the proposed reference, as they lead to a lower minimum supply voltage than regular-$V_{T}$ (RVT) devices and ensure the proper operation of the 2T voltage reference at low temperature, as will be discussed in Section~\ref{subsec:implementation_in_scaled_technologies_A}. Then, we choose to invert the topology shown in Fig.~\ref{fig:2_family_picture}(b), i.e., nMOS and pMOS devices are swapped, as well as ground and supply connections. This change makes it easier to reach a low reference current and avoids using several pwell voltages, requiring deep trench isolation (DTI) in PDSOI or triple-well devices in bulk. Finally, the sizing of the SCM and its current mirror is similar to \cite{CamachoGaleano_2005, CamachoGaleano_2008}, and performed at 25$^\circ$C. It consists of the following steps:
\begin{enumerate}
	\item Compute $V_{REF}$ using (\ref{eq:vref_PTAT});
	\item Determine the inversion level of $M_7$ by solving (\ref{eq:vref_if7}) for $i_{f7}$. By definition, the inversion level of $M_6$ is computed as $i_{f6} = \alpha i_{f7}$, and the sensitivity $S_{I_{REF}}$ is calculated using (\ref{eq:Siref_PTAT});
	\item Compute the aspect ratio and width of transistors $M_6$ and $M_7$, forming the SCM. $S_7$ is computed as
	\begin{equation}
		S_7 = \frac{N I_{REF}}{I_{SQ7} i_{f7}} \label{eq:S7}
	\end{equation}
	based on (\ref{eq:id_ACM}), while $S_6$ is calculated from (\ref{eq:S6_over_S7});
	\item Compute the aspect ratio and width of $M_{3-5}$ using (\ref{eq:id_ACM}).
\end{enumerate}
\begin{figure}[!t]
	\centering
	\includegraphics[width=.435\textwidth]{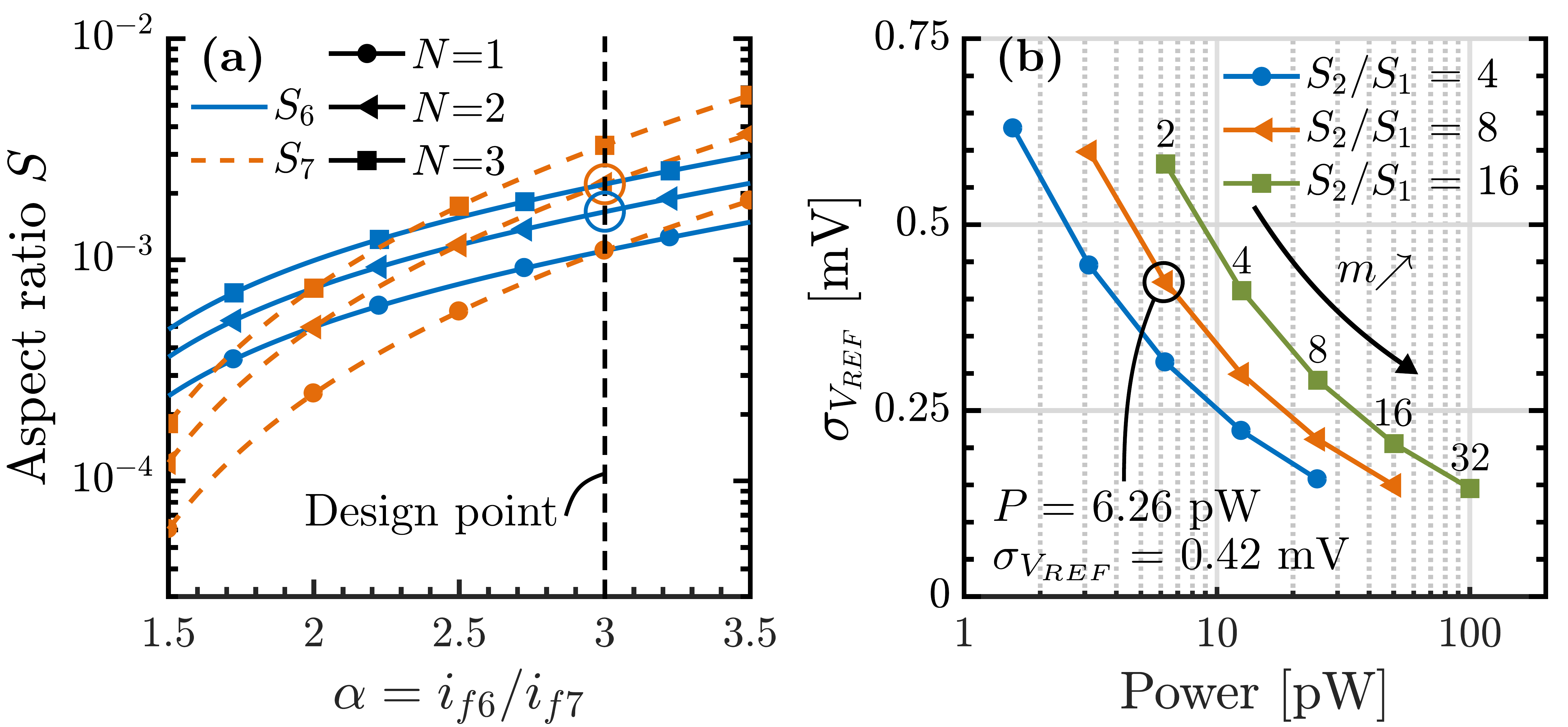}
	\caption{(a) Aspect ratio of transistors $M_{6-7}$ as a function of $\alpha$, for different values of $N$, and for $S_2/S_1 = 8$, assuming the use of LVT pMOS devices for the SCM. (b) Trade-off between reference voltage mismatch, obtained from $10^4$ Monte-Carlo runs, and power consumption in nominal conditions (TT, 1.2 V, 25$^\circ$C), as a function of $m$ and for different values of $S_2/S_1$. The unitary device is an LVT pMOS with a 1-$\mu$m width and 20-$\mu$m length.}
	\label{fig:5_ptat_vref_design}
\end{figure}
\begin{table}[!t]
\centering
\caption{Sizing of the nA-range PTAT current reference in XFAB 0.18-$\mu$m PDSOI.}
\label{table:sizing_PTAT}
\begin{scriptsize}
\begin{tabular}{lcccccc}
	\toprule
	& & \multicolumn{3}{c}{Sizing algorithm output} & \multicolumn{2}{c}{Final implementation}\\
	\cmidrule(l){3-5} \cmidrule(l){6-7}
	& Type & $W$ [$\mu$m] & $L$ [$\mu$m] & $i_f$ & $W$ [$\mu$m] & $L$ [$\mu$m]\\
	\midrule
	$M_1$ & LVT pMOS & 4$\times$1 & 20 & - & 4$\times$1 & 20\\
	$M_2$ & LVT pMOS & 32$\times$1 & 20 & - & 32$\times$1 & 20\\
	$M_3$ & LVT pMOS & 0.34 & 10 & 0.12 & 0.35 & 10\\
	$M_4$ & LVT nMOS & 1.03 & 2$\times$25 & 0.05 & 2$\times$0.5 & 2$\times$25\\
	$M_5$ & LVT nMOS & 2.06 & 2$\times$25 & 0.05 & 4$\times$0.5 & 2$\times$25\\
	$M_6$ & LVT pMOS & 0.82 & 20$\times$25 & 11.41 & \textbf{0.64} & 20$\times$25\\
	$M_7$ & LVT pMOS & 1.10 & 20$\times$25 & 3.80 & \textbf{0.86} & 20$\times$25\\
	\bottomrule
\end{tabular}
\end{scriptsize}
\end{table}
This methodology is implemented in Matlab, and the transistor properties, summarized in Table~\ref{table:transistor_properties}, are computed from DC SPICE simulations following \cite{Jespers_2017}. Contrary to \cite{CamachoGaleano_2005, CamachoGaleano_2008}, we do not limit ourselves to the methodology detailed hereabove, but we provide guidelines on how to select the main design parameters required by this methodology, namely $S_2/S_1$, $\alpha$ and $N$. Steps~1) and 2) are mostly technology-agnostic, as only the subthreshold slope factor $n$ is required at this stage. Indeed, the two other parameters coming into play are $S_2/S_1$ and $\alpha$, which do not depend on the technology choice. Fig.~\ref{fig:4_ptat_design}(a) depicts sensitivity $S_{I_{REF}}$ as a function of these two parameters, and allows to select a couple $(S_2/S_1;\:\alpha)$ close to the target value for $S_{I_{REF}}$, here arbitrarily set to 5~$\%$/mV. The chosen design point ($S_2/S_1 = 8;\:\alpha = 3$) is marked by a red star in Fig.~\ref{fig:4_ptat_design}(a) and yields a sensitivity of 5.14~$\%$/mV. Furthermore, the trends observed in Fig.~\ref{fig:4_ptat_design}(a) are better understood by looking at Figs.~\ref{fig:4_ptat_design}(b) and (c). Increasing $S_2/S_1$ pushes $M_{6-7}$ into moderate inversion, as evidenced by the increase of $i_{f7}$ [Fig.~\ref{fig:4_ptat_design}(b)], due to the higher reference voltage $V_{REF}$ applied to the SCM [Fig.~\ref{fig:4_ptat_design}(c)]. It should be noted that, for a fixed $\alpha$, the sensitivity approximately improves as $1/i_{f7}$, as indicated by the first term in (\ref{eq:Siref_PTAT}). Then, decreasing $\alpha$ provides a second tuning knob for improving $S_{I_{REF}}$, by pushing $M_7$ even further into moderate inversion [Fig.~\ref{fig:4_ptat_design}(b)]. Next, the results of step~3) are technology-dependent, because they rely on (i) the specific sheet current (Table~\ref{table:transistor_properties}) and (ii) the $I_{REF}$ target, here set to 0.1~nA. Fig.~\ref{fig:5_ptat_vref_design}(a) represents the aspect ratio of $M_{6-7}$ as a function of $\alpha$, for different values of $N$ and for a fixed voltage reference corresponding to $S_2/S_1 = 8$. Similarly to Fig.~\ref{fig:4_ptat_design}(b), decreasing $\alpha$ pushes $M_{6-7}$ into moderate inversion, thus decreasing their aspect ratio at constant current. Besides, higher values of the multiplicative factor $N$ increase $S_{6-7}$, as expected from (\ref{eq:S6_over_S7}) and (\ref{eq:S7}), facilitating the implementation of these transistors at the cost of a higher power consumption. Here, we select $N = 2$, which gives $S_6 = 1.65 \times 10^{-3}$ and $S_7 = 2.20 \times 10^{-3}$. The very low values obtained for these aspect ratios cannot be achieved with a single device, but will be implemented as a composite transistor, i.e., the series connection of several devices. Step~4) is straightforward as it is a direct consequence of previous choices.\\
\begin{table}[!t]
\centering
\caption{Resistor properties in the XFAB 0.18-$\mu$m PDSOI technology. Density is measured at 25$^\circ$C and TC is characterized over the -40-to-125$^\circ$C temperature range. All the resistors have a width of 2~$\mu$m and a length of 10~$\mu$m.}
\label{table:resistor_properties}
\resizebox{.475\textwidth}{!}{%
\begin{tabular}{lcccccc}
	\toprule
	\multirow{3}{*}{Resistor type} & \multicolumn{3}{c}{N+} & \multicolumn{3}{c}{P+}\\
	\cmidrule(l){2-4} \cmidrule(l){5-7}
	& diff. & poly. & poly. & diff. & poly. & poly.\\
	& & & (high res.) & & & (high res.)\\
	\midrule
	Density $[\Omega/\square]$ & 65 & 339 & 6564 & 80 & 295 & 1058\\
	TCR $[$ppm/$^\circ$C$]$ & 1388 & 1303 & 3562 & 4037 & 102 & 802\\
	\bottomrule
\end{tabular}%
}
\end{table}
\begin{figure}[!t]
	\centering
	\includegraphics[width=.403\textwidth]{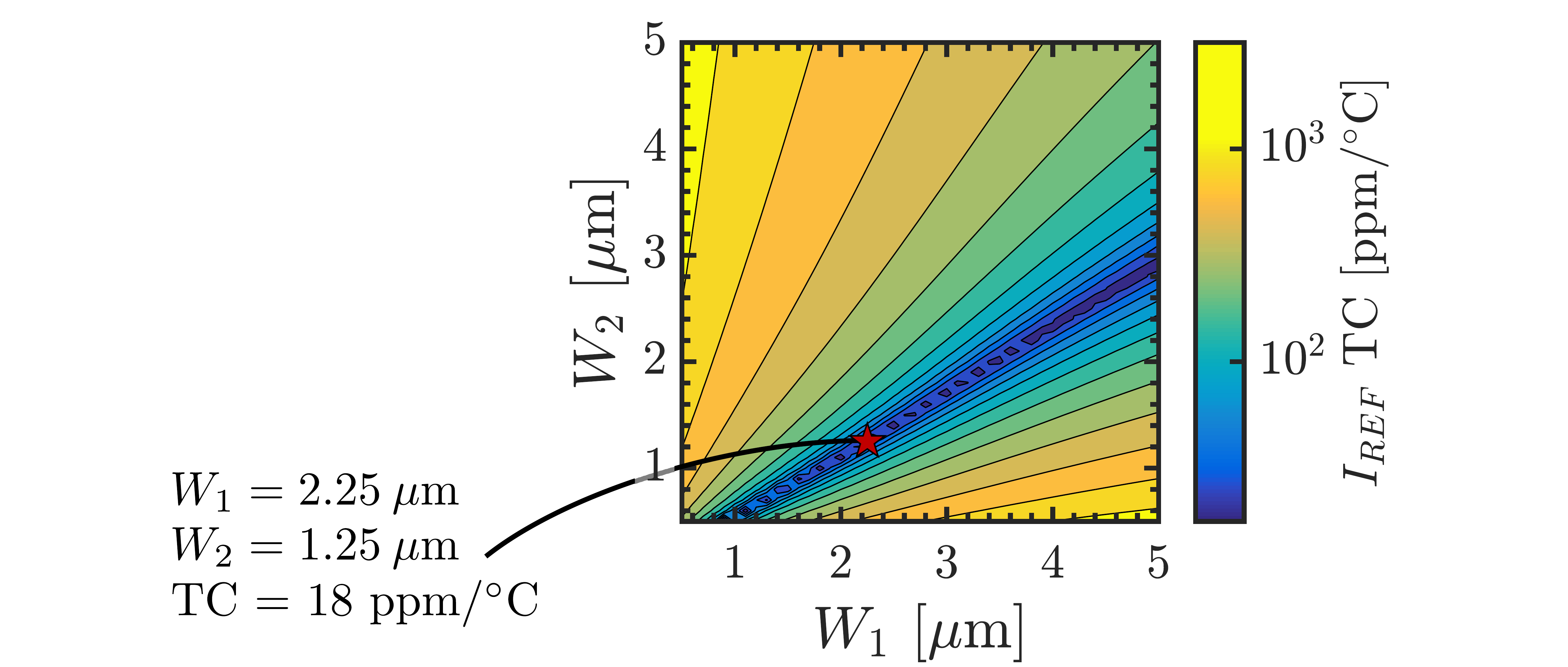}
	\caption{Current reference TC in the -40-to-125$^\circ$C range, as a function of transistor widths $W_1$ and $W_2$. The red star indicates the chosen design point. TC is computed as explained in Section~\ref{sec:simulation_and_measurement_results}.}
	\label{fig:6_cwt_design_iref_TC_vs_W1W2}
\end{figure}
\indent Finally, let us consider the sizing of the 2T voltage reference. As mentioned earlier, LVT pMOS devices are selected to ensure a proper operation in all process corners down to -40$^\circ$C, as will be explained in Section~\ref{subsec:implementation_in_scaled_technologies_A}. In addition, a length of 20~$\mu$m is chosen to reduce the output conductance of $M_2$ and improve the LS, as captured by (\ref{eq:LS_iref}). $M_{1-2}$ are then sized based on Fig.~\ref{fig:5_ptat_vref_design}(b), illustrating the trade-off between the standard deviation of the reference voltage due to local mismatch, denoted as $\sigma_{V_{REF}}$, and its power in nominal conditions. First, $S_2/S_1$ marginally impacts mismatch, as it is dominated by the threshold voltage variations of the smallest device, here $M_1$, whose sizes do not change with $S_2/S_1$. Consequently, increasing $S_2/S_1$ only results in a linear increase of the power consumption, shifting the curve to the right of Fig.~\ref{fig:5_ptat_vref_design}(b). Second, the multiplier $m$ improves mismatch as $1/\sqrt{m}$, following Pelgrom's law, while increasing power as $m$. A multiplier $m = 4$ is selected, in addition to the previously chosen $S_2/S_1 = 8$, and gives a power of 6.26~pW and a standard deviation of 0.42~mV.\\
\indent The transistor sizes output by the sizing algorithm are summarized in Table~\ref{table:sizing_PTAT}. Two main changes are applied for the final implementation: (i) $M_{4-5}$ are split into parallel devices to enable a common centroid layout, and (ii) $M_{6-7}$ are narrowed based on simulation results to match the 0.1-nA $I_{REF}$ target in nominal conditions. It should be noted that $\alpha$, and therefore $S_6/S_7$, must remain constant during this upscaling, to preserve the reference behavior.
\begin{figure}[!t]
	\centering
	\includegraphics[width=.455\textwidth]
	{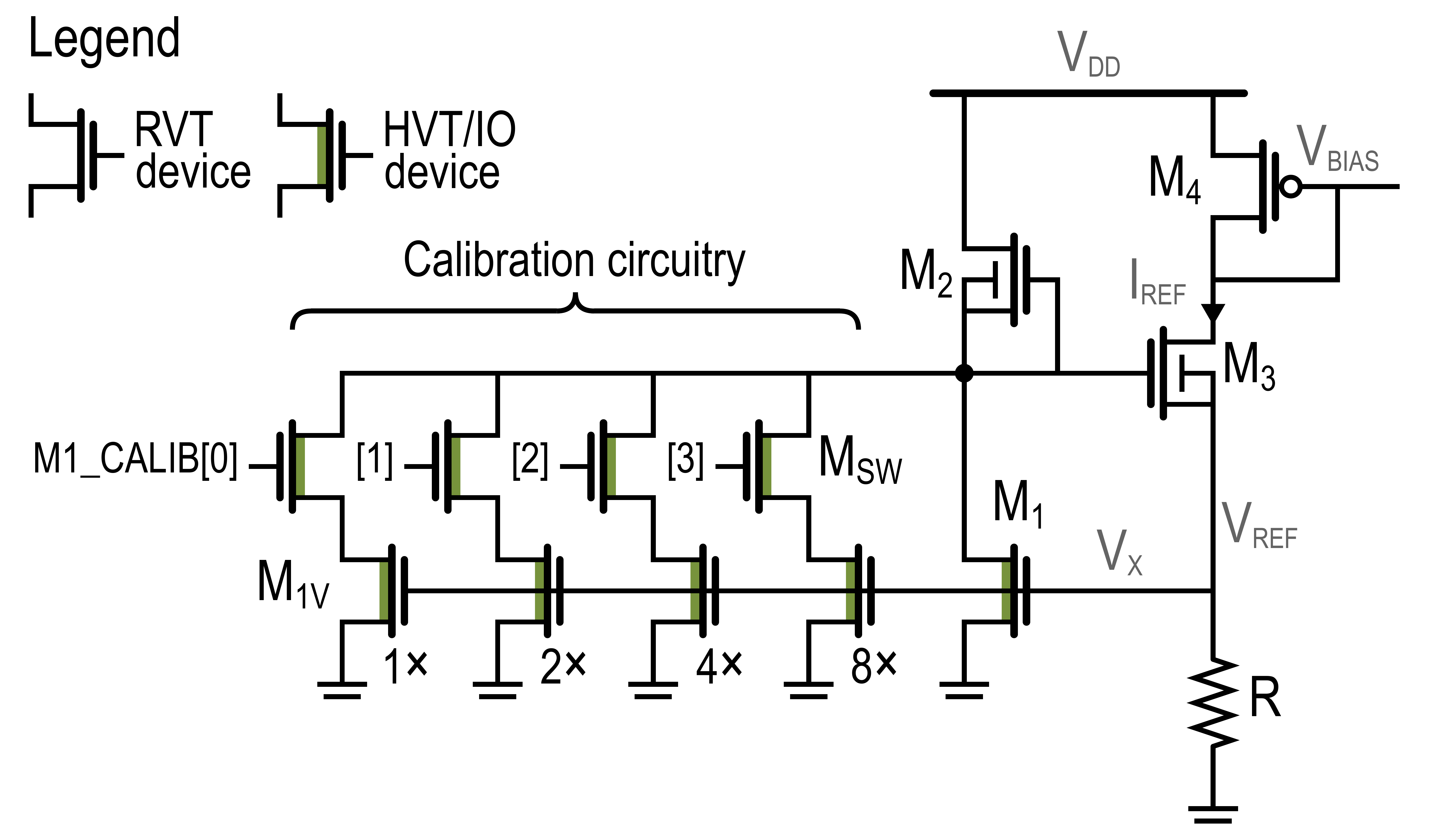}
	\caption{A 4-bit TC calibration scheme is implemented by changing the width of $M_1$ and thus the ratio $W_2/W_1$.}
	\label{fig:7_calibration_scheme}
\end{figure}

\subsection{$\mu$A-Range CWT Current Reference}
\label{subsec:circuit_design_and_sizing_methodology_B}
The sizing of the CWT current reference shown in Fig.~\ref{fig:2_family_picture}(a) focuses on minimizing the TC of $I_{REF}$ while achieving the current reference target, here set to 1~$\mu$A. The sizing methodology boils down to three steps:
\begin{enumerate}
	\item Select a resistor with a low temperature coefficient of resistance (TCR);
	\item Size transistors $M_{1-2}$ to minimize the TC of $I_{REF}$;
	\item Tune $R$ to reach the current reference target in TT.
\end{enumerate}
Note that, similarly to the PTAT current reference, we choose to invert the topology depicted in Fig.~\ref{fig:2_family_picture}(a). First, at step~1), selecting a resistor with a low TCR is desirable to avoid the 2T voltage reference from compensating large and highly nonlinear resistance variations, which would lead to a poor TC of $I_{REF}$. Table~\ref{table:resistor_properties} summarizes the properties of the available resistors, namely their density and their TCR, computed as
\begin{equation}
\textrm{TCR} = \frac{\left(R_{max} - R_{min}\right)}{R(25^\circ \textrm{C}) \left(T_{max}-T_{min}\right)}.
\end{equation}
P+ poly resistors, i.e., poly over a p-type-doped substrate, present the lowest TCR values, with 102 and 802~ppm/$^\circ$C for the regular and high resistance flavors, respectively. A regular P+ poly resistor (\texttt{rpp1}) is selected for its low TCR, despite its modest density of 295~$\Omega/\square$. Next, the objective of step~2) is to size $M_{1-2}$ to minimize the TC of $I_{REF}$. For the 2T voltage reference to operate properly, (\ref{eq:vref}) highlights the need for $M_1$ to have a larger threshold voltage than $M_2$, which is why an RVT and LVT pMOS are chosen. The length of both transistors being fixed to 5~$\mu$m to reach a competitive LS, Fig.~\ref{fig:6_cwt_design_iref_TC_vs_W1W2} demonstrates that the TC of $I_{REF}$ is minimized by a $W_2/W_1$ ratio of 0.56, as suggested by (\ref{eq:S2_over_S1_iref_CWT}). Widths of 2.25 and 1.25~$\mu$m are selected for $M_1$ and $M_2$, on the grounds of variability, giving a TC of 18~ppm/$^\circ$C. Finally, step~3) simply consists in tuning the value of the resistance to reach the 1-$\mu$A target. Transistors $M_{3-4}$ can be easily sized, with the main limitations being that in all PVT corners, $M_3$ must ensure that $M_1$ remains saturated, and $V_{SG4}$ must be larger than $4U_T$ for the current mirror to operate properly.\\
\indent Nevertheless, the ratio $W_2/W_1$ leading to the minimum TC is process-dependent. A calibration mechanism, represented in Fig.~\ref{fig:7_calibration_scheme}, is consequently implemented by tuning the width of $M_1$ with a 4-bit control signal. This allows to change $W_2/W_1$ from 0.37 to 0.83, leading to a TC for $I_{REF}$ in the 10-to-30-ppm/$^\circ$C range in most process corners, as will be shown in Section~\ref{subsec:simulation_and_measurement_results_B}. To conclude, the chosen transistor sizes for the CWT current reference without and with calibration are summed up in Table~\ref{table:sizing_CWT}.

\section{Simulation and Measurement Results}
\label{sec:simulation_and_measurement_results}
\begin{table}[!t]
\centering
\caption{Sizing of the $\mu$A-range CWT current reference, without and with TC calibration, in XFAB 0.18-$\mu$m PDSOI.}
\label{table:sizing_CWT}
\begin{scriptsize}
\begin{tabular}{lccccc}
	\toprule
	& & \multicolumn{2}{c}{w/o TC calib.} & \multicolumn{2}{c}{w/ TC calib.}\\
	& Type & $W$ [$\mu$m] & $L$ [$\mu$m] & $W$ [$\mu$m] & $L$ [$\mu$m]\\
	\midrule
	$M_1$ & RVT pMOS & 4$\times$2.25 & 5 & 12$\times$0.8 & 5\\
	$M_{1V}$ & RVT pMOS & / & / & 1 - 8$\times$0.8 & 5\\
	$M_{SW}$ & RVT pMOS & / & / & 0.22 & 5\\
	$M_2$ & LVT pMOS & 4$\times$1.25 & 5 & 4$\times$2 & 5\\
	$M_3$ & LVT pMOS & 10$\times$10 & 1 & 10$\times$10 & 1\\
	$M_4$ & LVT nMOS & 10$\times$10 & 1 & 10$\times$10 & 1\\
	$R$ & P+ poly. & 0.45 & 26$\times$13.1 & 0.45 & 26$\times$11.9\\
	\bottomrule
\end{tabular}
\end{scriptsize}
\end{table}
\begin{figure}[!t]
	\centering
	\includegraphics[width=.375\textwidth]{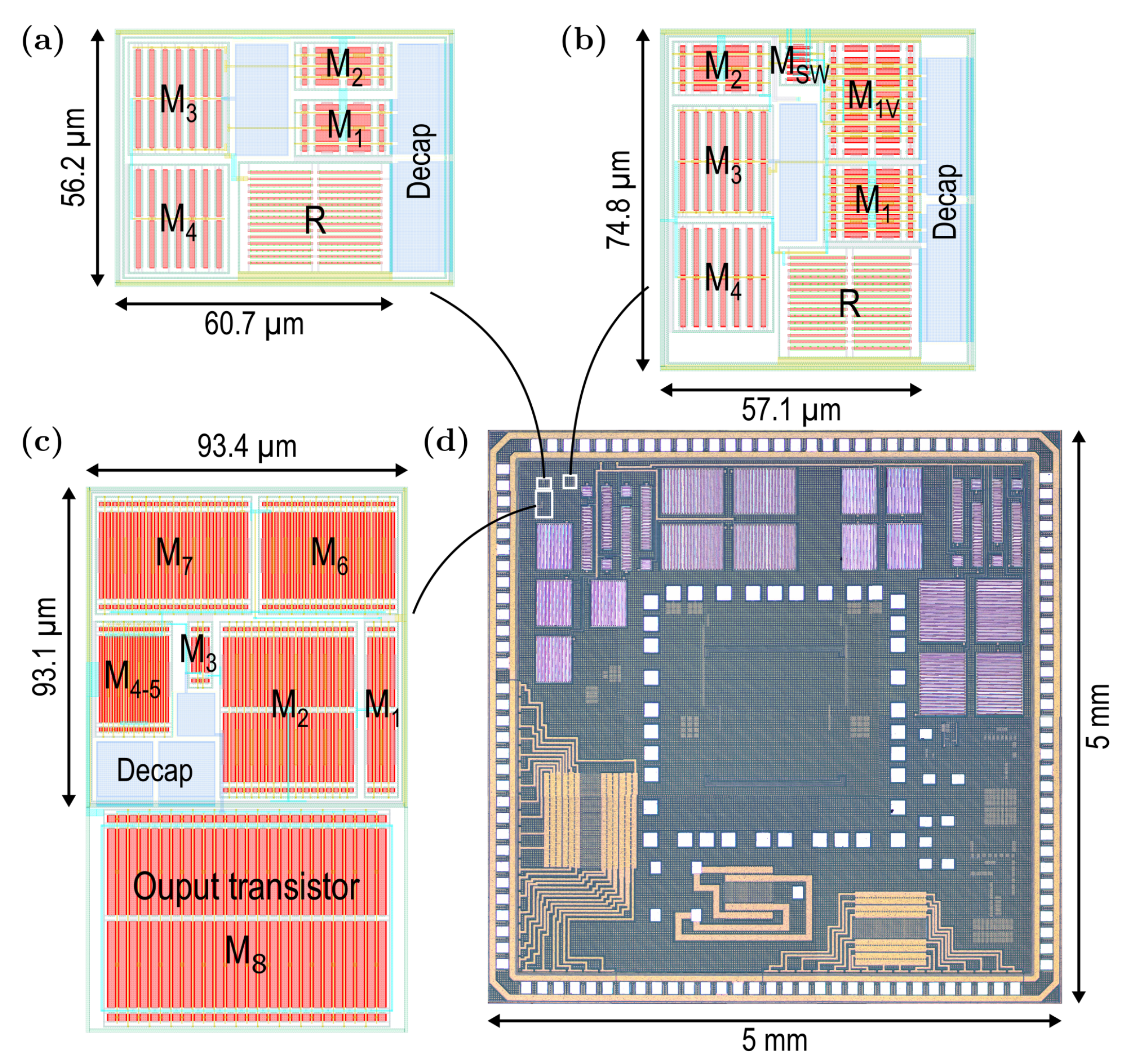}
	\caption{Chip microphotograph and layout of the current references in XFAB \mbox{0.18-$\mu$m} PDSOI: $\mu$A-range CWT current reference (a) without and (b) with calibration, (c) nA-range PTAT current reference and (d) 25-mm$^2$ 0.18-$\mu$m PDSOI chip microphotograph.}
	\label{fig:8_chip_microphotograph}
\end{figure}
This section presents the simulation and measurement results for the three current references fabricated in XFAB \mbox{0.18-$\mu$m} PDSOI, which behaves like a conventional bulk technology because it does not suffer from floating-body effects. Their layouts are shown in Fig.~\ref{fig:8_chip_microphotograph}, together with the chip microphotograph. Simulations are performed post-layout, to account for non-idealities due to layout effects and parasitic diodes. Current measurements are carried out with a Keithley 2636A source meter, connected to the PCB including the chip through triaxial cables. The PCB is placed in an Espec SH-261 climatic chamber, in which temperature is swept from -40 to 85$^\circ$C by steps of 5$^\circ$C while leaving humidity uncontrolled. Temperature is limited to 85$^\circ$C as some pieces of equipment cannot withstand higher temperatures.

In what follows, the LS and TC are computed using the box method, i.e.,
\begin{IEEEeqnarray}{CCL}
	\textrm{LS} & = & \frac{\left(I_{REF,max}-I_{REF,min}\right)}{I_{REF,avg} \left(V_{DD,max}-V_{DD,min}\right)} \times 100 \: \%\textrm{/V,}\IEEEnonumber\\
	\textrm{TC} & = & \frac{\left(I_{REF,max}-I_{REF,min}\right)}{I_{REF,avg} \left(T_{max}-T_{min}\right)} \times 10^6 \: \textrm{ppm/}^\circ\textrm{C,}\IEEEnonumber%
\end{IEEEeqnarray}
where $I_{REF,min}$, $I_{REF,avg}$ and $I_{REF,max}$ respectively stand for the minimum, average and maximum reference current among the considered range. $V_{DD,min}$ (resp. $T_{min}$) and $V_{DD,max}$ (resp. $T_{max}$) refer to the lower and upper bounds of the voltage (resp. temperature) range.

\subsection{nA-Range PTAT Current Reference}
\label{subsec:simulation_and_measurement_results_A}
\begin{figure}[!t]
	\centering
	\includegraphics[width=.424\textwidth]{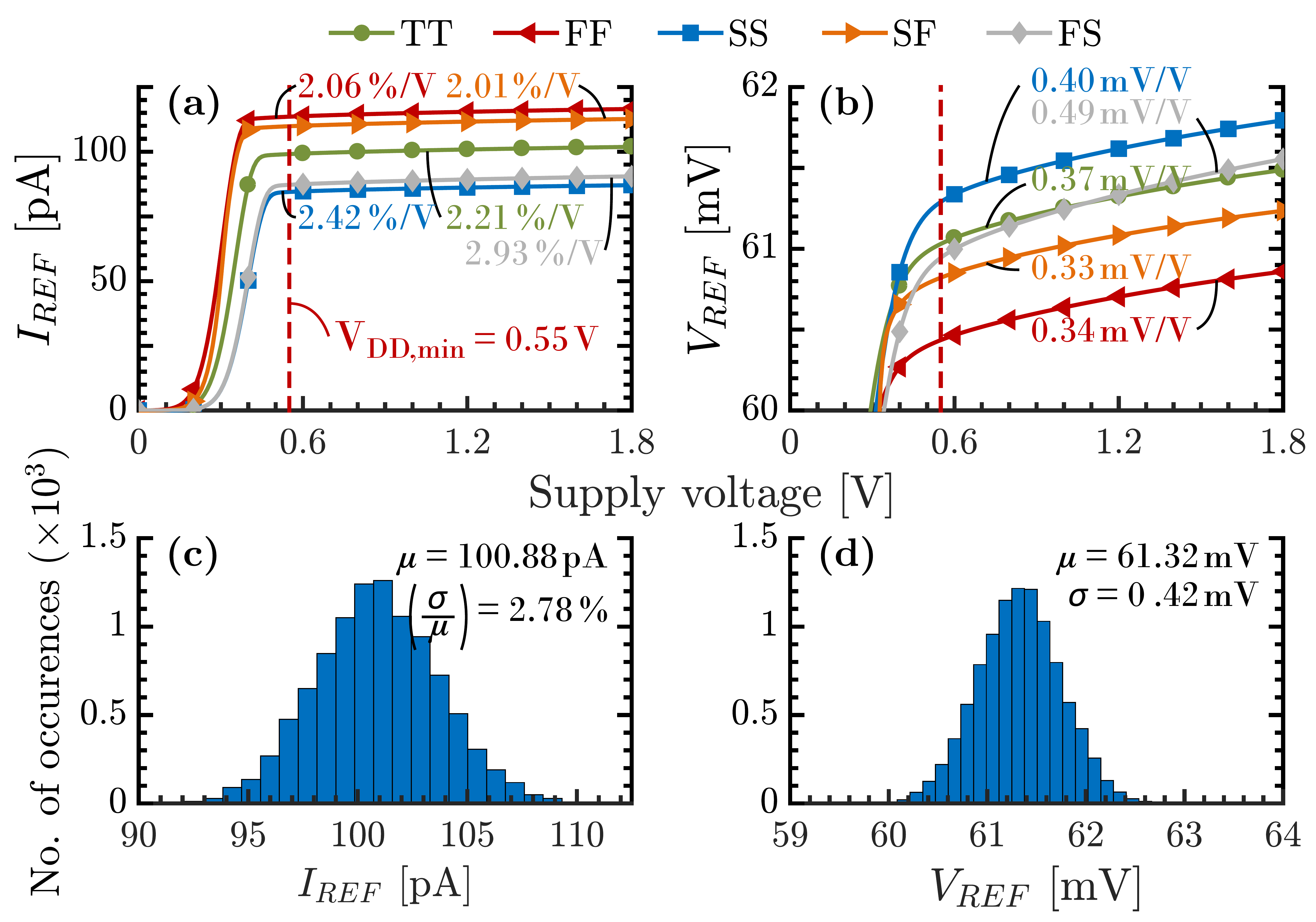}
	\caption{Simulated nA-range PTAT current reference (a) $I_{REF}$ and (b) $V_{REF}$ as a function of supply voltage, in all process corners and at 25$^\circ$C. Histograms of (c) $I_{REF}$ and (d) $V_{REF}$ for 10$^4$ Monte-Carlo runs in nominal conditions, i.e., in TT, at 1.2~V and 25$^\circ$C.}
	\label{fig:9_sim_xfab180_iptat_vs_vdd_mc}
\end{figure}
\begin{figure}[!t]
	\centering
	\includegraphics[width=.477\textwidth]{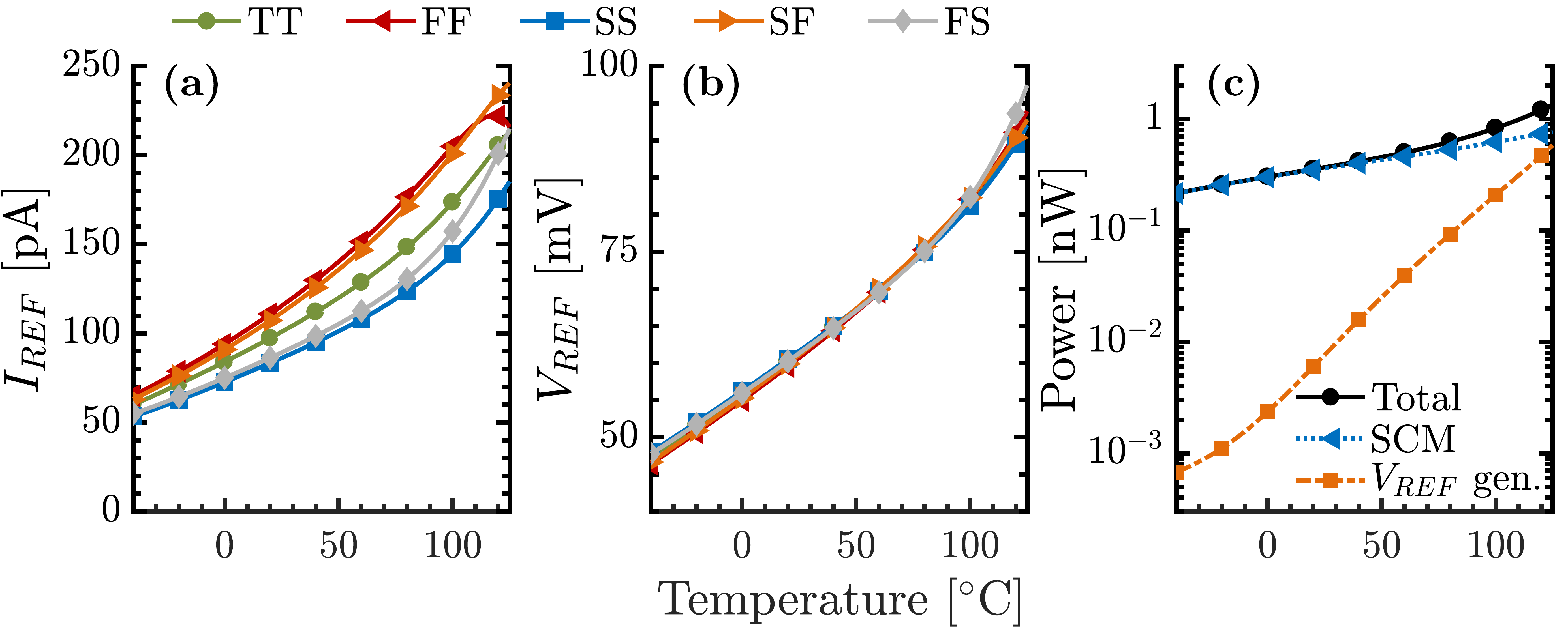}
	\caption{Simulated temperature dependence of (a) $I_{REF}$ and (b) $V_{REF}$ in all process corners and at 1.2~V. (c) Power breakdown between the SCM and the 2T voltage reference in TT, as a function of temperature.}
	\label{fig:10_sim_xfab180_iptat_vs_T}
\end{figure}
\begin{figure}[!t]
	\centering
	\includegraphics[width=.424\textwidth]{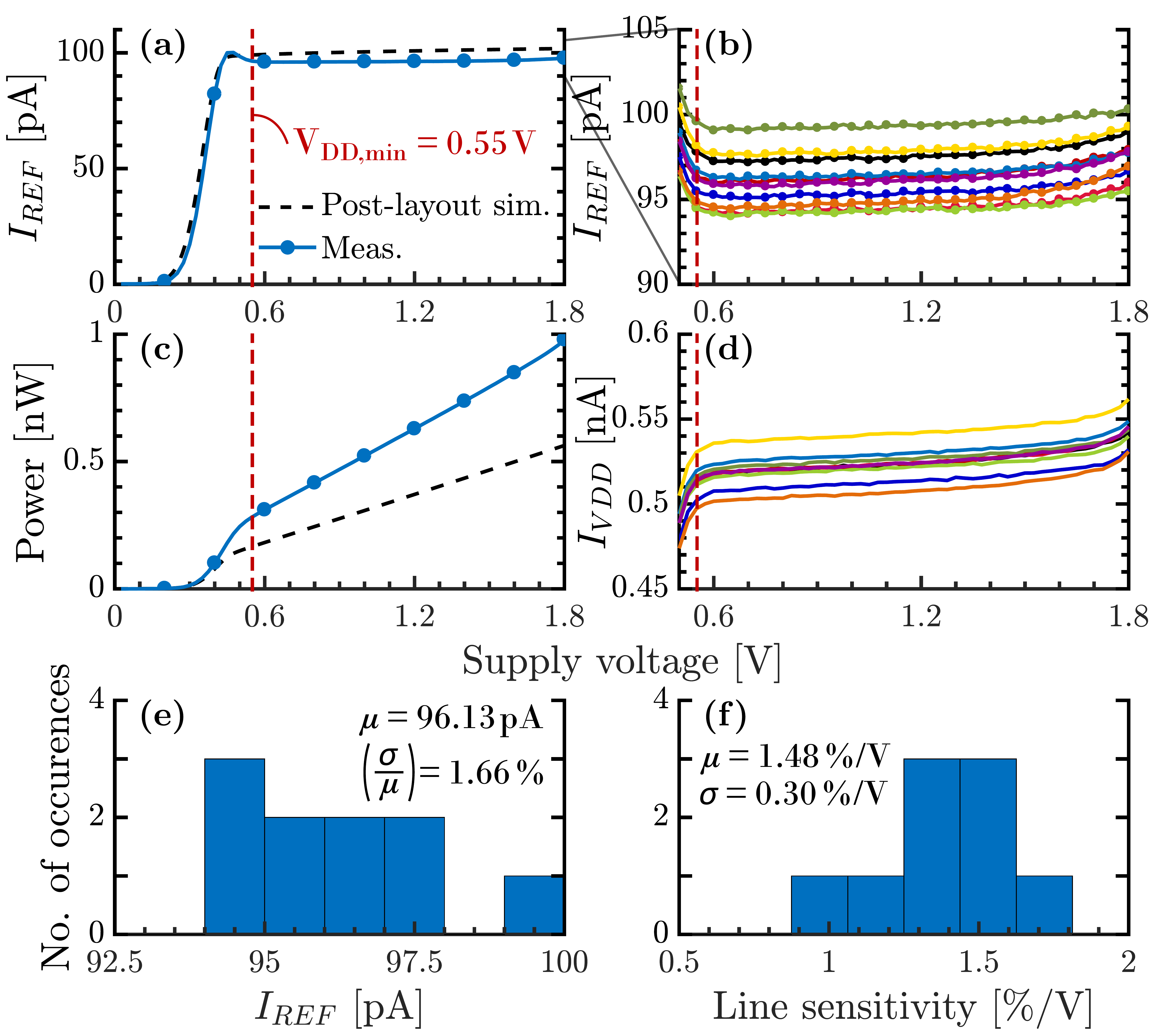}
	\caption{Measured average (a) $I_{REF}$ and (c) power consumption vs. $V_{DD}$ at 25$^\circ$C, with (b)-(d) details of the 10 dies, for the nA-range current reference. Histograms of (e) $I_{REF}$ and (f) LS across the 10 dies.}
	\label{fig:11_meas_iptat_vs_vdd}
\end{figure}
\begin{figure}[!t]
	\centering
	\includegraphics[width=.424\textwidth]{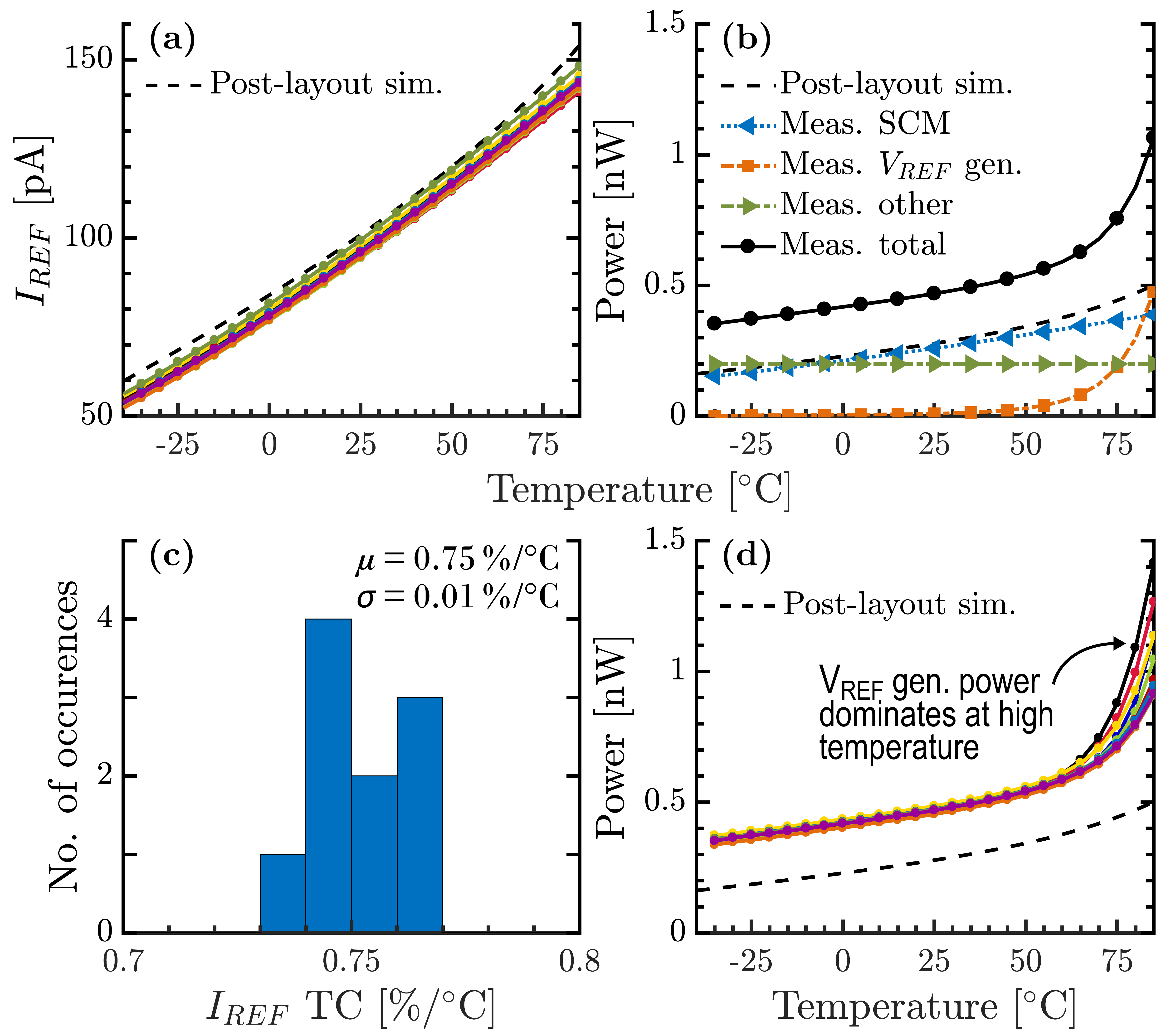}
	\caption{Measured temperature dependence of (a) $I_{REF}$ and (d) power consumption at 0.9~V, for all 10 dies. (b) Power breakdown between the SCM and the 2T voltage reference. (d) Histogram of TC across the 10 dies.}
	\label{fig:12_meas_iptat_vs_T}
\end{figure}
Post-layout simulation results are represented in Figs.~\ref{fig:9_sim_xfab180_iptat_vs_vdd_mc} and \ref{fig:10_sim_xfab180_iptat_vs_T}. First, as discussed in Section~\ref{subsec:a_family_of_current_references_A}, the LS can be predicted by (\ref{eq:LS_iref}). With $S_{I_{REF}}=$ 5.14~$\%$/mV given by the design point and the LS of $V_{REF}$ equal to 0.37~mV/V, as shown in Fig.~\ref{fig:9_sim_xfab180_iptat_vs_vdd_mc}(b), (\ref{eq:LS_iref}) predicts an LS of \mbox{1.90~$\%$/V}. Figs.~\ref{fig:9_sim_xfab180_iptat_vs_vdd_mc}(a) and (b) display that $V_{DD,min}$ is around 0.55~V and is limited by the minimum voltage required to bias the 2T voltage reference. The LS is 2.21~$\%$/V from 0.55 to 1.8~V in TT, which is larger than the \mbox{1.90-$\%$/V} prediction as it does not account for the supply voltage dependence coming from the current mirror biasing the SCM. In addition, LS is relatively stable among process corners and is mainly impacted by process variations of $I_{REF}$, i.e., +14.5~$\%$ in FF and -11.6~$\%$ in SS. Thus, the worst-case LS is 2.93~$\%$/V in FS. Next, the variability of $I_{REF}$ can be linked to $\sigma_{V_{REF}}=$ 0.42~mV through (\ref{eq:MC_iref}), giving a prediction of 2.16~$\%$. Again, this value is lower than the simulation result of 2.78~$\%$ exhibited in Figs.~\ref{fig:9_sim_xfab180_iptat_vs_vdd_mc}(c) and (d), as the local mismatch in the SCM and the current mirror has not been accounted for. Finally, Fig.~\ref{fig:10_sim_xfab180_iptat_vs_T} depicts the temperature dependence of the PTAT current reference. Fig.~\ref{fig:10_sim_xfab180_iptat_vs_T}(a) highlights that $I_{REF}$ is approximetaly linear with temperature up to 100$^\circ$C in TT. Above it, nonlinearities induced by leakage in parasitic nwell/psub diodes appear. This translates to a \mbox{0.92-$\%/^\circ$C} TC for $I_{REF}$ between -40 and 125$^\circ$C in TT. Fig.~\ref{fig:10_sim_xfab180_iptat_vs_T}(b) confirms the PTAT behavior of $V_{REF}$. Fig.~\ref{fig:10_sim_xfab180_iptat_vs_T}(c) demonstrates that power consumption is linear with temperature below 60$^\circ$C, and is dominated by the SCM which draws a current proportional to $I_{REF}$. Then, power increases exponentially with temperature due to the 2T voltage reference, whose power consumption scales with drain-to-source leakage. Finally, the startup time in typical conditions is 238 ms [Fig.~\ref{fig:19_xfab180_startup}(a)].\\
\begin{figure}[!t]
	\centering
	\includegraphics[width=.424\textwidth]{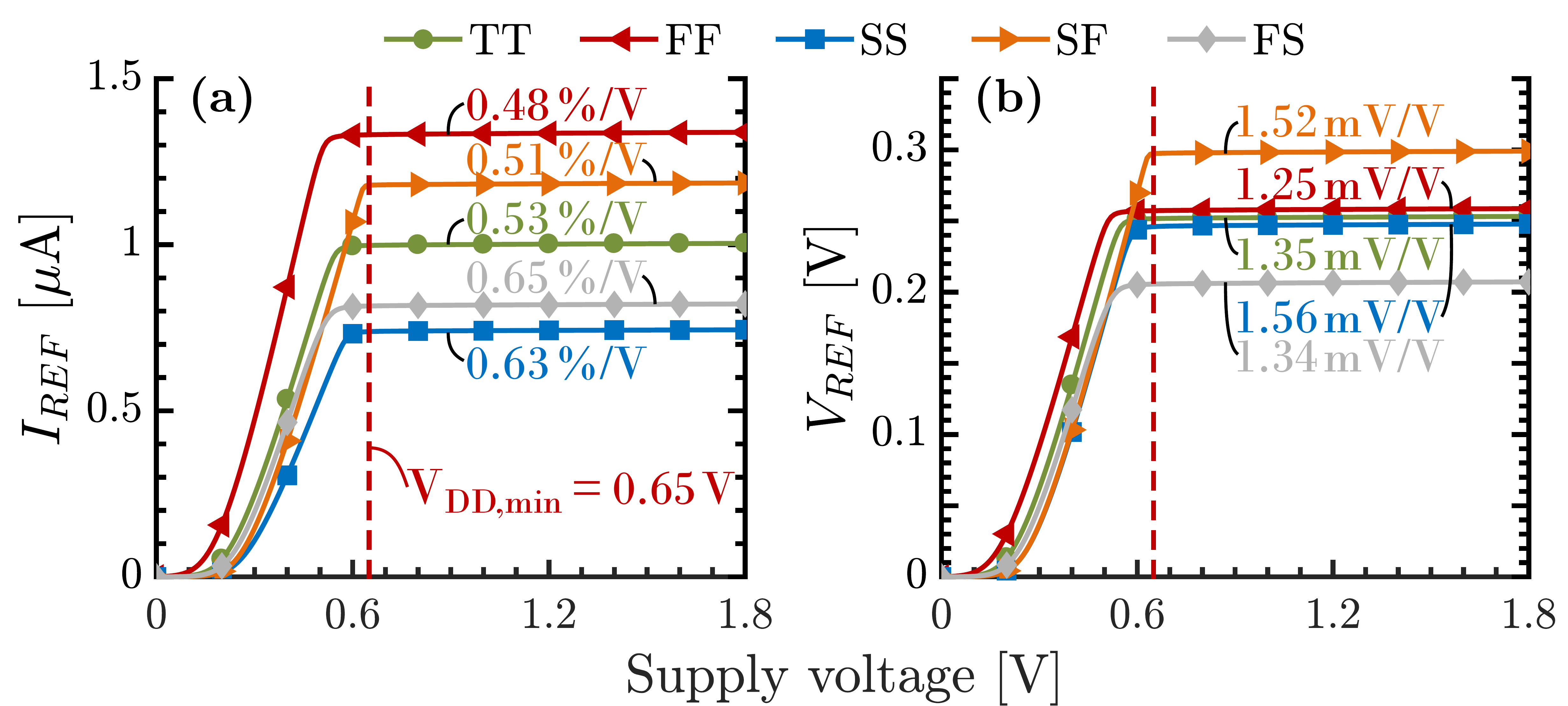}
	\caption{Simulated $\mu$A-range CWT current reference (a) $I_{REF}$ and (b) $V_{REF}$, as a function of supply voltage, in all process corners and at 25$^\circ$C.}
	\label{fig:13_sim_xfab180_icwt_vs_vdd}
\end{figure}
\begin{figure}[!t]
	\centering
	\includegraphics[width=.477\textwidth]{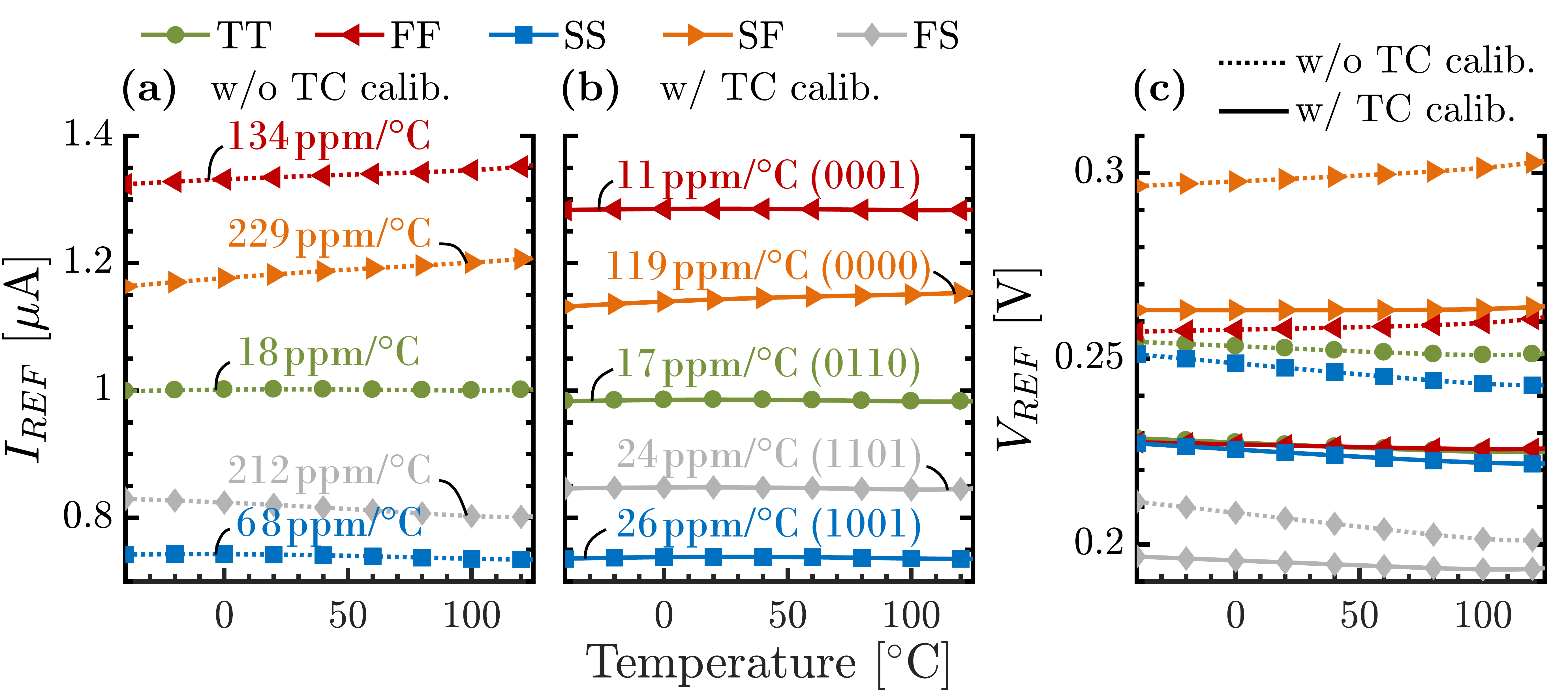}
	\caption{Simulated temperature dependence of $I_{REF}$ (a) without and (b) with TC calibration, and (c) $V_{REF}$, in all process corners and at 1.2~V.}
	\label{fig:14_sim_xfab180_icwt_vs_T}
\end{figure}
\begin{figure}[!t]
	\centering
	\includegraphics[width=.424\textwidth]{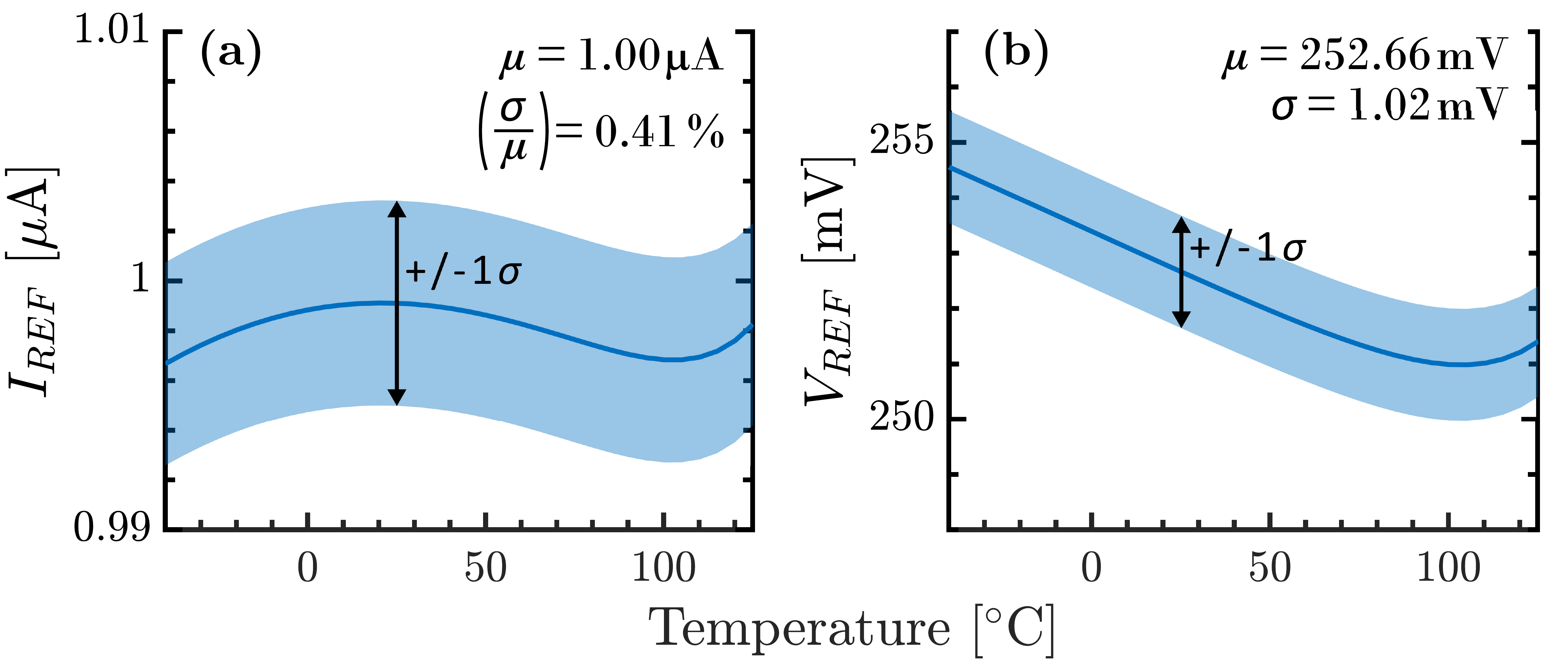}
	\caption{Simulated temperature dependence of (a) $I_{REF}$ and (b) $V_{REF}$ with $\pm 1\sigma$, for 10$^4$ Monte-Carlo runs in TT and at 1.2~V.}
	\label{fig:15_sim_xfab180_icwt_mc}
\end{figure}
\indent Measurement results are presented in Figs.~\ref{fig:11_meas_iptat_vs_vdd} and \ref{fig:12_meas_iptat_vs_T}. In what follows, the reference current is measured as the drain current of $M_8$ [Fig.~\ref{fig:8_chip_microphotograph}(c)], divided by a fixed ratio of 36.4, obtained from nominal simulations. Firstly, Fig.~\ref{fig:11_meas_iptat_vs_vdd}(a) shows the average $I_{REF}$ as a function of $V_{DD}$, and only differs from post-layout simulations by a small overshoot at 0.5~V. Regarding power consumption [Fig.~\ref{fig:11_meas_iptat_vs_vdd}(c)], the measured value is larger than the simulated one by roughly 2$\times$ due to additional leakage in the 2T voltage reference, but a minimum power of 0.28~nW is reached at 0.55~V. Moreover, details of the 10 measured dies are provided in Figs.~\ref{fig:11_meas_iptat_vs_vdd}(b) and (d). Then, the variability of $I_{REF}$ across the 10 dies is depicted in Fig.~\ref{fig:11_meas_iptat_vs_vdd}(e), with an average and $(\sigma/\mu)$ equal to 0.096~nA and 1.66~$\%$, compared to 0.101~nA and 2.78~$\%$ in simulation. Given that in Fig.~\ref{fig:11_meas_iptat_vs_vdd}(f), the average LS (1.48~$\%$/V) is also lower than in simulation (2.21~$\%$/V), we hypothesize that both observations are explained by a lower-than-expected sensitivity $S_{I_{REF}}$. Finally, Fig.~\ref{fig:12_meas_iptat_vs_T}(a) illustrates the temperature dependence of $I_{REF}$ for all 10 dies, which closely matches the post-layout simulations. The measured average TC is 0.75~$\%$/$^\circ$C from -40 to 85$^\circ$C [Fig.~\ref{fig:12_meas_iptat_vs_T}(c)], with a standard deviation of 0.01~$\%$/$^\circ$C. Regarding power consumption, Fig.~\ref{fig:12_meas_iptat_vs_T}(b) highlights that the increase from simulation to measurement comes from a temperature-independent current, which could possibly originate from the 2T voltage reference. Furthermore, we observe in Fig.~\ref{fig:12_meas_iptat_vs_T}(d) that the 2T voltage reference power becomes dominant above 60$^\circ$C, and leads to a larger growth than in TT simulations.

\subsection{$\mu$A-Range CWT Current Reference}
\label{subsec:simulation_and_measurement_results_B}
Post-layout simulation results are shown in Figs.~\ref{fig:13_sim_xfab180_icwt_vs_vdd} to \ref{fig:15_sim_xfab180_icwt_mc}. First, Figs.~\ref{fig:13_sim_xfab180_icwt_vs_vdd}(a) and (b) depict $I_{REF}$ and $V_{REF}$ as a function of $V_{DD}$. Similarly to the PTAT reference, they highlight that $V_{DD,min}$ is limited by the 2T voltage reference and is equal to 0.65~V. From (\ref{eq:LS_iref}), we know that LS is inversely proportional to the intrinsic gain $g_m/g_d$. Given the large $g_m/g_d$ in this technology \cite{Murmann_2006}, an LS of 0.53~$\%$/V is reached in TT from 0.65 to 1.8~V, with the worst-case LS corresponding to 0.65~$\%$/V in FS. Besides, $I_{REF}$ is strongly impacted by process variations, with +33.3~$\%$ in FF and -25.6~$\%$ in SS. Furthermore, the temperature dependence of $I_{REF}$ and $V_{REF}$ is represented in Fig.~\ref{fig:14_sim_xfab180_icwt_vs_T} without and with TC calibration. Fig.~\ref{fig:14_sim_xfab180_icwt_vs_T}(a) illustrates that without TC calibration, an excellent TC of 18~ppm/$^\circ$C is obtained in TT, but it degrades in skewed corners with 229~ppm/$^\circ$C PTAT in SF and 212~ppm/$^\circ$C CTAT in FS, due to a change of temperature dependence of $V_{REF}$ [Fig.~\ref{fig:14_sim_xfab180_icwt_vs_T}(c)]. Nevertheless, Fig.~\ref{fig:14_sim_xfab180_icwt_vs_T}(b) demonstrates that a simple 4-bit calibration of the $W_2/W_1$ width ratio [Fig.~\ref{fig:7_calibration_scheme}] can reduce the TC below 30~ppm/$^\circ$C in all process corners except for SF, in which the 119~ppm/$^\circ$C PTAT TC would require a slightly larger tuning range. Lastly, in Fig.~\ref{fig:15_sim_xfab180_icwt_mc}(a), $I_{REF}$ presents a 2$^{\mathrm{nd}}$ order temperature dependence below 100$^\circ$C, and leakage leads to a current increase above this limit. Moreover, the ($\sigma/\mu$)'s of $I_{REF}$ and $V_{REF}$ are equal, as stated by (\ref{eq:MC_iref}), and amount to 0.41~$\%$ [Figs.~\ref{fig:15_sim_xfab180_icwt_mc}(a) and (b)], featuring a perfect match between theory and simulations. Regarding $I_{REF}$, its average value is 1~$\mu$A at 25$^\circ$C, and its average TC reaches 14.9~ppm/$^\circ$C, with a variability of 2.14~$\%$. Lastly, the startup time is 5 ms in typical conditions [Fig.~\ref{fig:19_xfab180_startup}(b)].\\
\begin{figure}[!t]
	\centering
	\includegraphics[width=.424\textwidth]{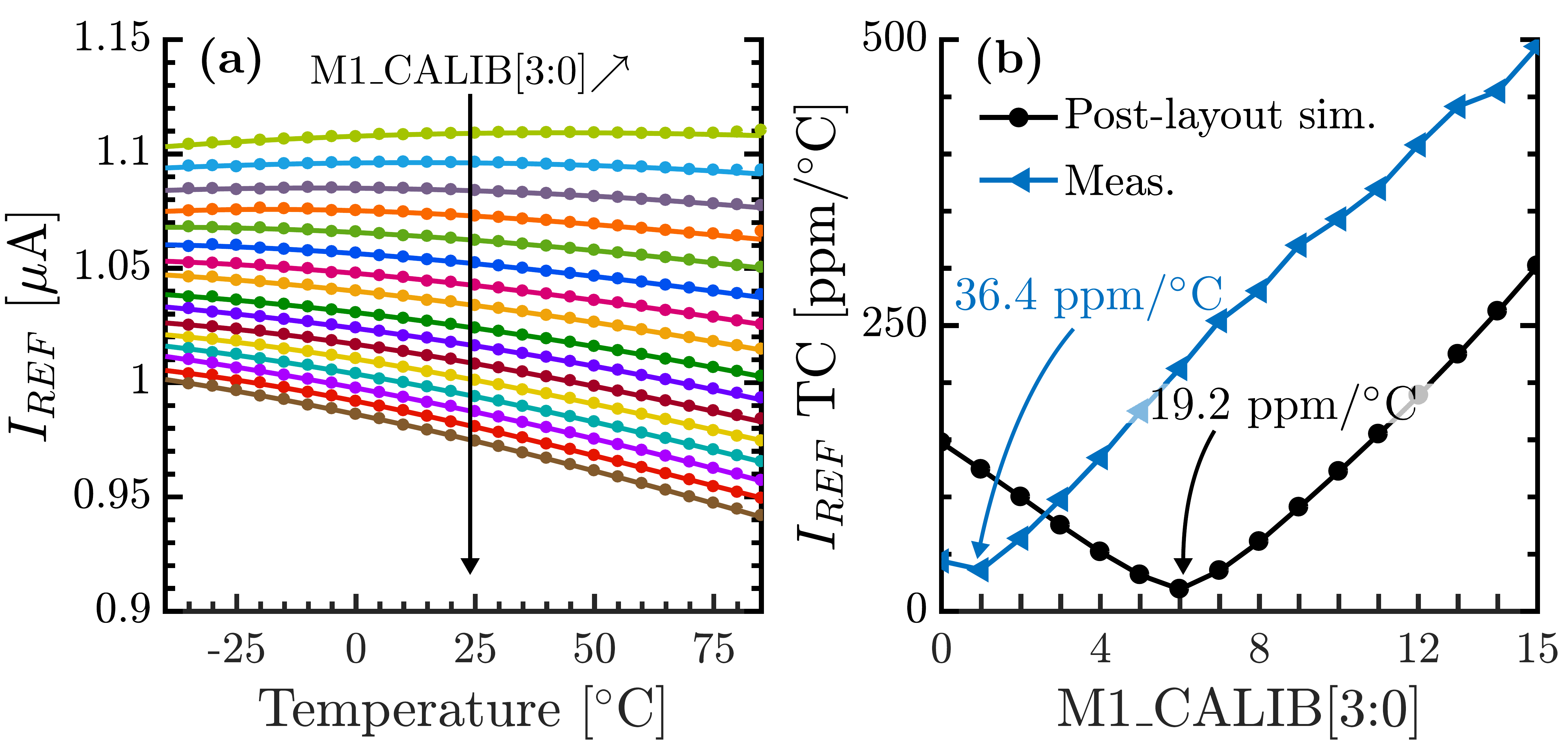}
	\caption{Measured (a) temperature dependence of $I_{REF}$ at 0.9~V, for all 16 calibration codes on a single die, for the $\mu$A-range current reference. (b) $I_{REF}$ TC obtained for each calibration code, in TT post-layout simulations and in measurement.}
	\label{fig:16_meas_icwt_vs_T_calib}
\end{figure}
\begin{figure}[!t]
	\centering
	\includegraphics[width=.466\textwidth]{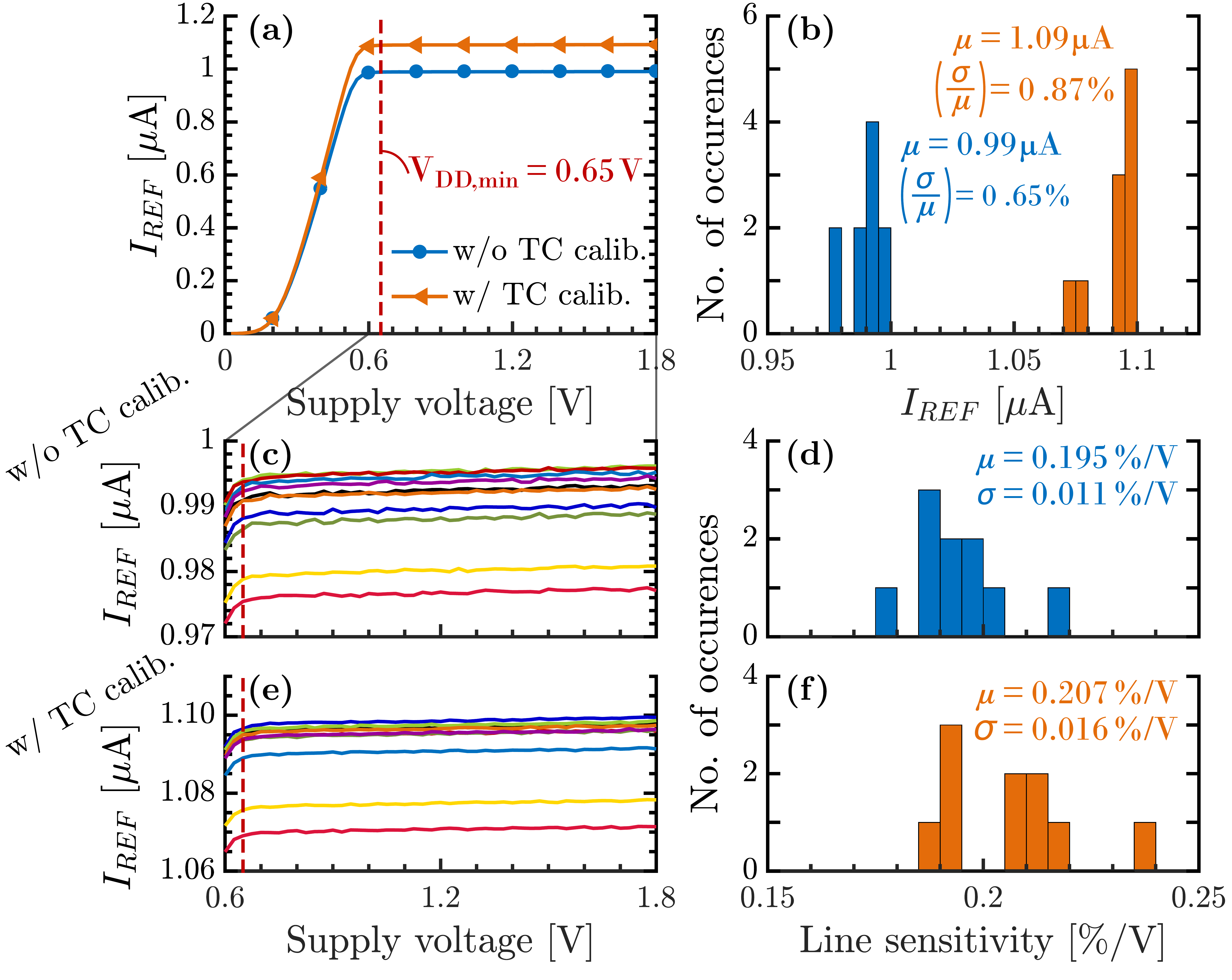}
	\caption{Measured (a) average $I_{REF}$ vs. $V_{DD}$, and details of the 10 dies (c) without and (e) with TC calibration, at 25$^\circ$C. Histograms of (b) $I_{REF}$ and LS (d) without and (f) with TC calibration, across the 10 dies.}
	\label{fig:17_meas_icwt_vs_vdd}
\end{figure}
\begin{figure}[!t]
	\centering
	\includegraphics[width=.424\textwidth]{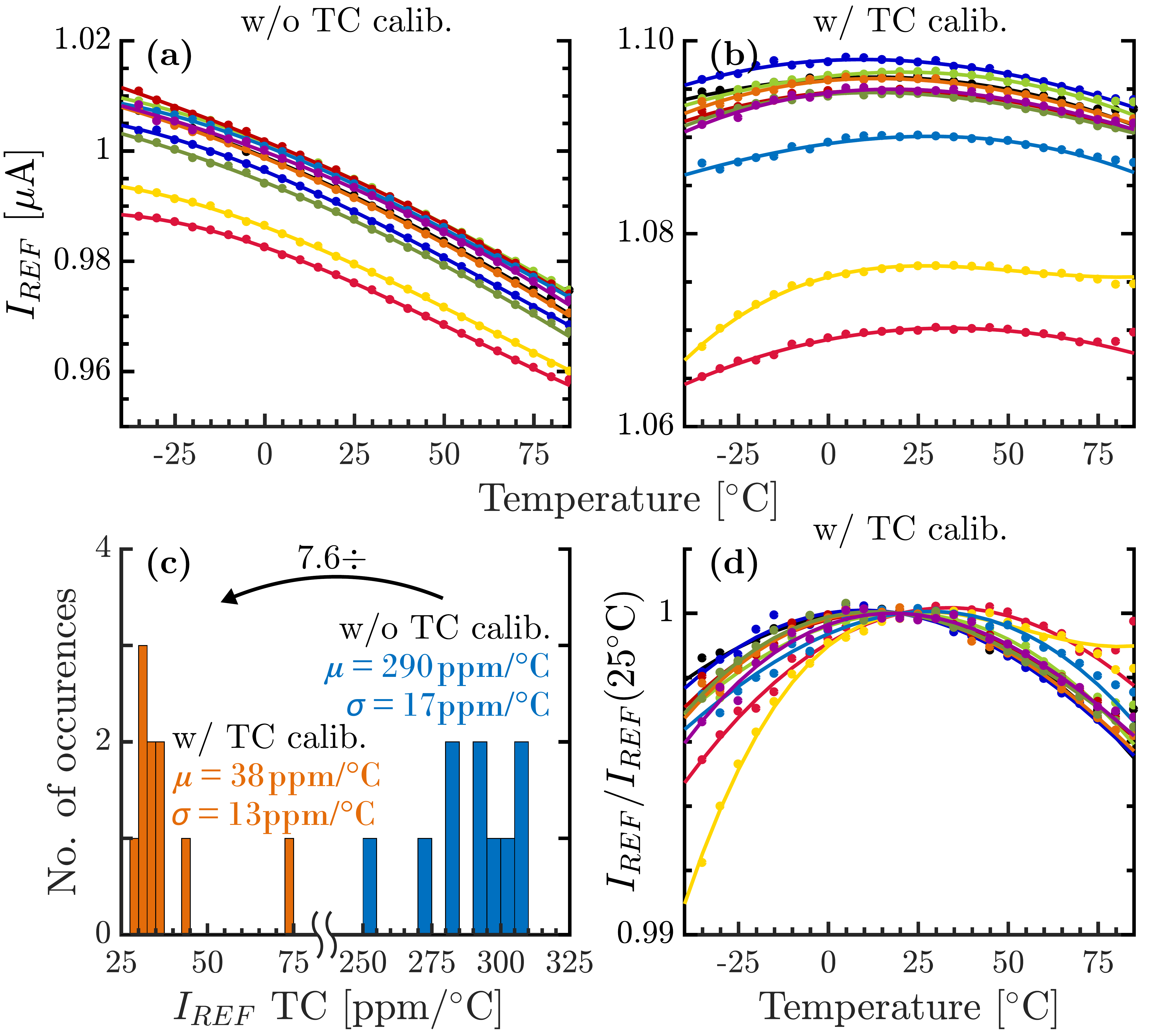}
	\caption{Measured temperature dependence of $I_{REF}$ (a) without TC calibration, (b) with TC calibration and (d) with TC calibration and normalization, at 0.9~V. (c) Histogram of $I_{REF}$ TC across the 10 dies.}
	\label{fig:18_meas_icwt_vs_T}
\end{figure}
\begin{figure}[!t]
	\centering
	\includegraphics[width=.424\textwidth]{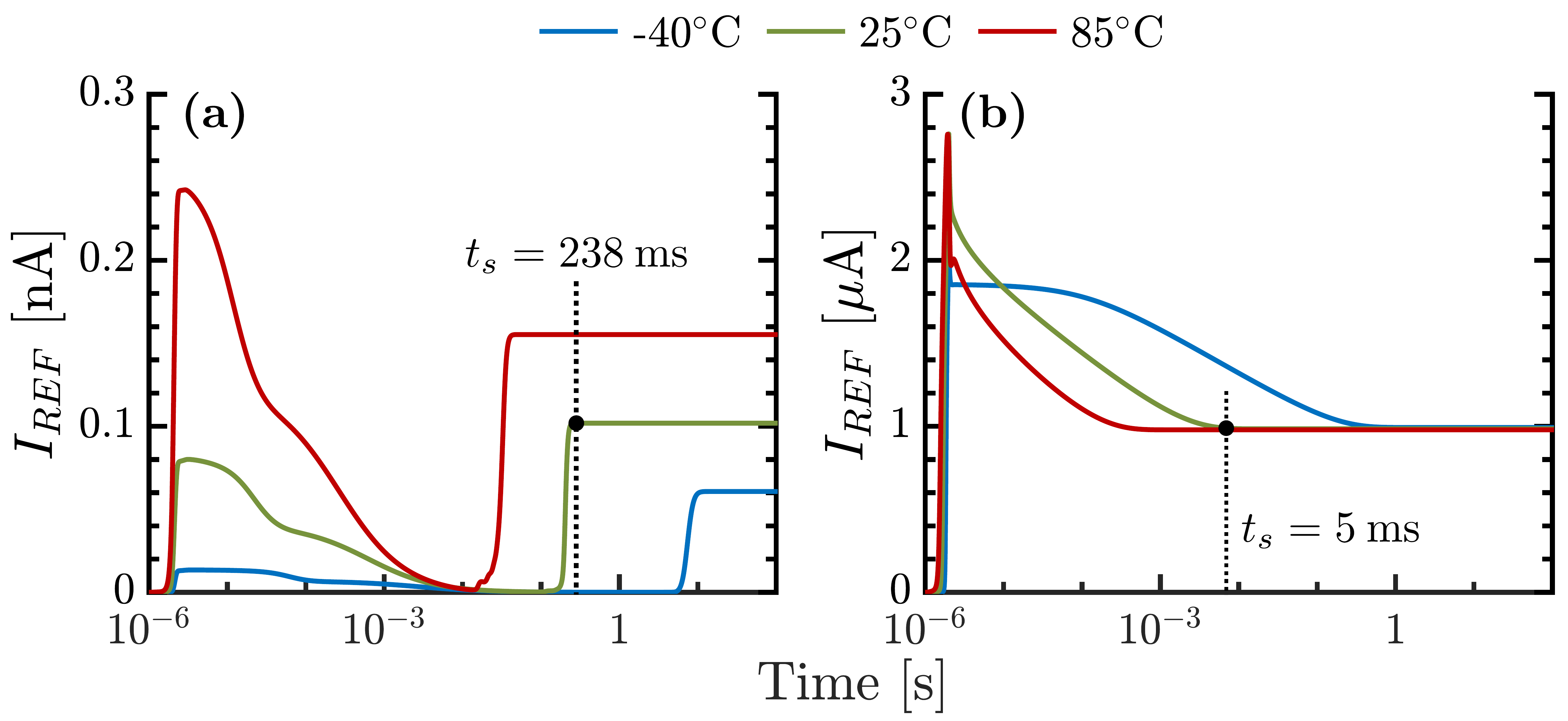}
	\caption{Simulated startup waveforms of the (a) PTAT nA-range and (b) CWT $\mu$A-range current references in the TT process, with a 1.8-V and 1-$\mu$s rise time supply voltage.}
	\label{fig:19_xfab180_startup}
\end{figure}
Measurement results are summarized in Figs.~\ref{fig:16_meas_icwt_vs_T_calib} to \ref{fig:18_meas_icwt_vs_T}. As the TC of $I_{REF}$ is mostly impacted by process variations rather than local mismatch, we select the optimal code by sweeping all 16 calibration codes on a single die from -40 to 85$^\circ$C, as shown in Fig.~\ref{fig:16_meas_icwt_vs_T_calib}(a). Alternatively, a two-point calibration at 25 and 85$^\circ$C would yield the same result. Because we are using a voltage reference with pMOS devices for $M_{1-2}$ and the control switches, increasing the calibration code decreases the actual $W_1$, thus reducing the PTAT part of $V_{REF}$ and boosting the CTAT behavior of $I_{REF}$. Based on Fig.~\ref{fig:16_meas_icwt_vs_T_calib}(b), we select a code of 0x0001 with a \mbox{36.4-ppm/$^\circ$C} TC, which is different from the simulated one, here 0x0110 with a \mbox{19.2-ppm/$^\circ$C} TC, likely due to process variations. Next, Fig.~\ref{fig:17_meas_icwt_vs_vdd}(a) shows the average $I_{REF}$ behavior with supply voltage, with a $V_{DD,min}$ at 0.65~V and a slight increase of $I_{REF}$ from 0.99 to 1.09~$\mu$A due to TC calibration [Fig.~\ref{fig:18_meas_icwt_vs_T}(b)]. In both cases, the average LS is around 0.20~$\%$/V, which represents a 2.7$\times$ reduction compared to simulations and could result from a larger $g_m/g_d$. Variability reaches 0.65~$\%$ and 0.87~$\%$ without and with TC calibration [Fig.~\ref{fig:17_meas_icwt_vs_vdd}(b)]. The larger value compared to 0.41~$\%$ in simulation likely comes from the yellow and red curves on the bottom of Figs.~\ref{fig:17_meas_icwt_vs_vdd}(c) and (e), which appear to be outliers leading to an overestimation of $(\sigma/\mu)$. To conclude, the temperature dependence of $I_{REF}$ is represented in Figs.~\ref{fig:18_meas_icwt_vs_T}(a) and (b). On one side, without calibration, $I_{REF}$ presents a CTAT behavior leading to a TC of 290~ppm/$^\circ$C in the \mbox{-40-to-85$^\circ$C} range. On the other side, with calibration, the TC reduces down to 38~ppm/$^\circ$C, corresponding to a 7.6$\times$ improvement. In Figs.~\ref{fig:18_meas_icwt_vs_T}(b) and (d), we observe a 2$^{\mathrm{nd}}$ order temperature dependence below 70$^\circ$C, as predicted by simulations. A small current surge above this limit comes from the power consumption of the 2T voltage reference, which is measured together with the reference current, but is not taken into account in the TC.

\section{Implementation in Scaled Technologies}
\label{sec:implementation_in_scaled_technologies}
Novel current references are often developed in the \mbox{0.18-$\mu$m} technology node or above. However, their portability to common scaled technologies such as \mbox{65-nm} bulk, \mbox{28-nm} fully-depleted SOI (FDSOI), or \mbox{14-nm} FinFET, is hardly discussed and poses a series of challenges. This section details these challenges and how they can be overcome for the proposed family of current references by focusing on the implementation of the 2T voltage reference.
\begin{figure}[!t]
	\centering
	\includegraphics[width=.424\textwidth]{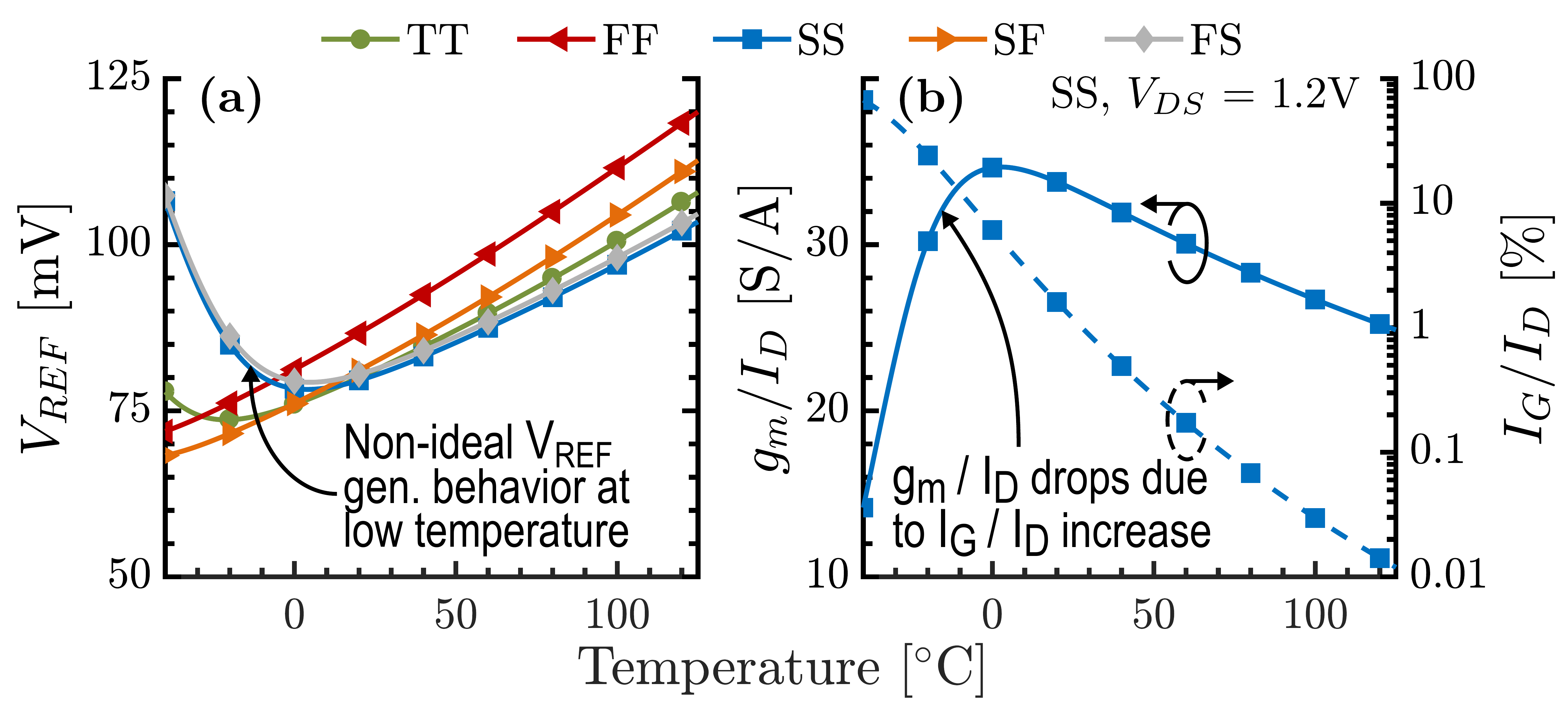}
	\caption{Temperature dependence of (a) a PTAT voltage reference implemented with pMOS devices in 65-nm bulk and (b) the $g_m/I_D$ and $I_G/I_D$ of a zero-$V_{GS}$ pMOS transistor in the SS process corner with $V_{DS} =$ 1.2~V.}
	\label{fig:20_vref_design_gmid_vs_T}
\end{figure}
\begin{figure}[!t]
	\centering
	\includegraphics[width=.424\textwidth]{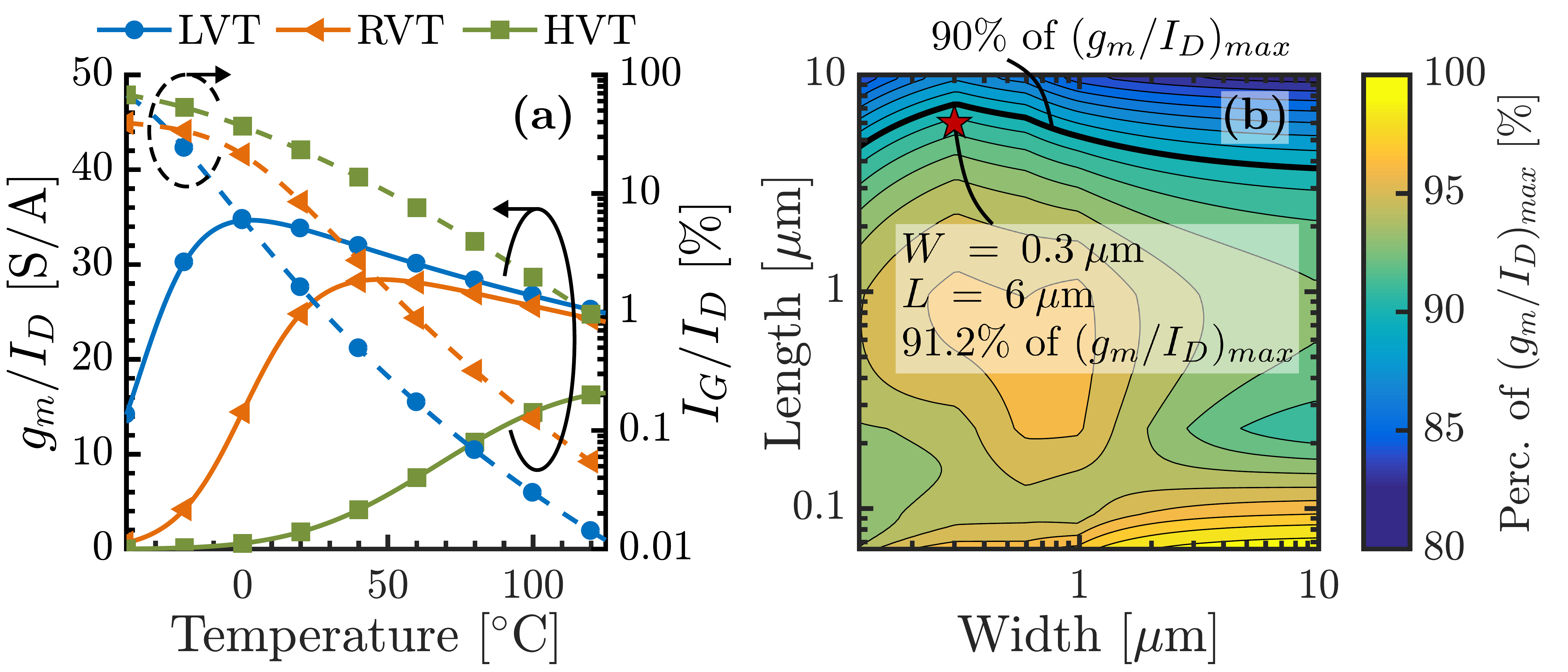}
	\caption{For a 65-nm zero-$V_{GS}$ pMOS in SS, (a) impact of the transistor flavor on the $g_m/I_D$ and $I_G/I_D$ curves vs. temperature, and (b) at 0$^\circ$C, fraction of $(g_m/I_D)_{max}$ obtained at zero $V_{GS}$ depending on the transistor sizes, with $(g_m/I_D)_{max}$ the maximum of the $g_m/I_D$ vs. $V_{GS}$ curve.}
	\label{fig:21_vref_design_transistor_type_dim}
\end{figure}

\subsection{Leakage-Induced Non-Idealities}
\label{subsec:implementation_in_scaled_technologies_A}
A PTAT voltage reference, implemented with pMOS devices in 65-nm bulk, is used to highlight the non-idealities of the 2T voltage reference. In Fig.~\ref{fig:20_vref_design_gmid_vs_T}(a), we note an increase of the reference voltage below 0$^\circ$C in the slow pMOS process corners (SS and FS). A similar yet weaker behavior can be observed below -20$^\circ$C in TT. Fig.~\ref{fig:20_vref_design_gmid_vs_T}(b) depicts the transconductance efficiency $g_m/I_D$ and the ratio between gate and drain currents $I_G/I_D$, for a zero-$V_{GS}$ pMOS in SS, as a function of temperature. The non-ideal behavior noticed in Fig.~\ref{fig:20_vref_design_gmid_vs_T}(a) coincides with a drop in $g_m/I_D$, which should increase at low temperature as the transconductance efficiency in deep subthreshold is proportional to $(g_m/I_D)_{max} = 1/(nU_T)$. As gate leakage has increased by several orders of magnitude from \mbox{0.18-$\mu$m} to \mbox{65-nm} \cite{Lewyn_2009, Bohannon_2011}, this drop can be explained by an increased $I_G/I_D$ ratio, coming from the fact that the current flowing in the 2T voltage reference is a drain-to-source leakage, decreasing exponentially with a temperature reduction, while the gate leakage remains approximately constant with temperature. $I_G$ thus becomes non-negligible, reaching 6.5~$\%$ of $I_D$ at 0$^\circ$C in SS. Two tuning knobs can be used to mitigate this non-ideal behavior: the transistor type and sizes. First, Fig.~\ref{fig:21_vref_design_transistor_type_dim}(a) illustrates the impact of the transistor type using the same kind of curves as Fig.~\ref{fig:20_vref_design_gmid_vs_T}(b). It points out that increasing the threshold voltage from LVT to high-$V_{T}$ (HVT) degrades the voltage reference behavior by shifting the point at which $g_m/I_D$ drops, corresponding to an $I_G/I_D$ around 1 to 5~$\%$, to higher temperature. LVT devices are thus selected to implement the voltage reference, to ensure functionality at low temperature at the cost of an increased power consumption at high temperature. Second, Fig.~\ref{fig:21_vref_design_transistor_type_dim}(b) depicts the evolution with the transistor sizes of the ratio between $g_m/I_D$ at zero $V_{GS}$ and $(g_m/I_D)_{max}$. A length increase degrades this ratio, as it linearly increases $I_G$ by expanding the gate area, while simultaneously decreasing $I_D$ as $1/L$. A width increase has a limited impact, because at first order, $I_G$ and $I_D$ both increase linearly with it. Nevertheless, Fig.~\ref{fig:21_vref_design_transistor_type_dim}(b) shows that 2$^\mathrm{nd}$ order effects also come into play. Based on this figure, we select a design point with a $g_m/I_D$ at zero $V_{GS}$ equal to 91.2~$\%$ of $(g_m/I_D)_{max}$, corresponding to $W = 0.3\:\mu$m and $L = 6\:\mu$m. We could select a shorter length to further mitigate the non-ideal behavior of the voltage reference, but only at the cost of a larger LS and power consumption.\\
\indent Moreover, we argue that the $g_m/I_D$ drop is a useful technology indicator for design as it captures various leakage sources degrading the voltage reference behavior. Indeed, in \mbox{65-nm} bulk, the drop can be explained by the impact of gate leakage. Yet, in \mbox{0.18-$\mu$m} PDSOI and \mbox{28-nm} FDSOI, gate-induced drain leakage (GIDL) prevails, as the gate leakage is limited by the use of thick oxide and high-$\kappa$ gates, respectively. Monitoring the $g_m/I_D$ drop allows to capture any such effects, regardless of their origin, and makes it possible to size the voltage reference without thoroughly investigating them.

\subsection{Line Sensitivity Enhancement Techniques}
\label{subsec:implementation_in_scaled_technologies_B}
\begin{figure}[!t]
	\centering
	\includegraphics[width=.45\textwidth]{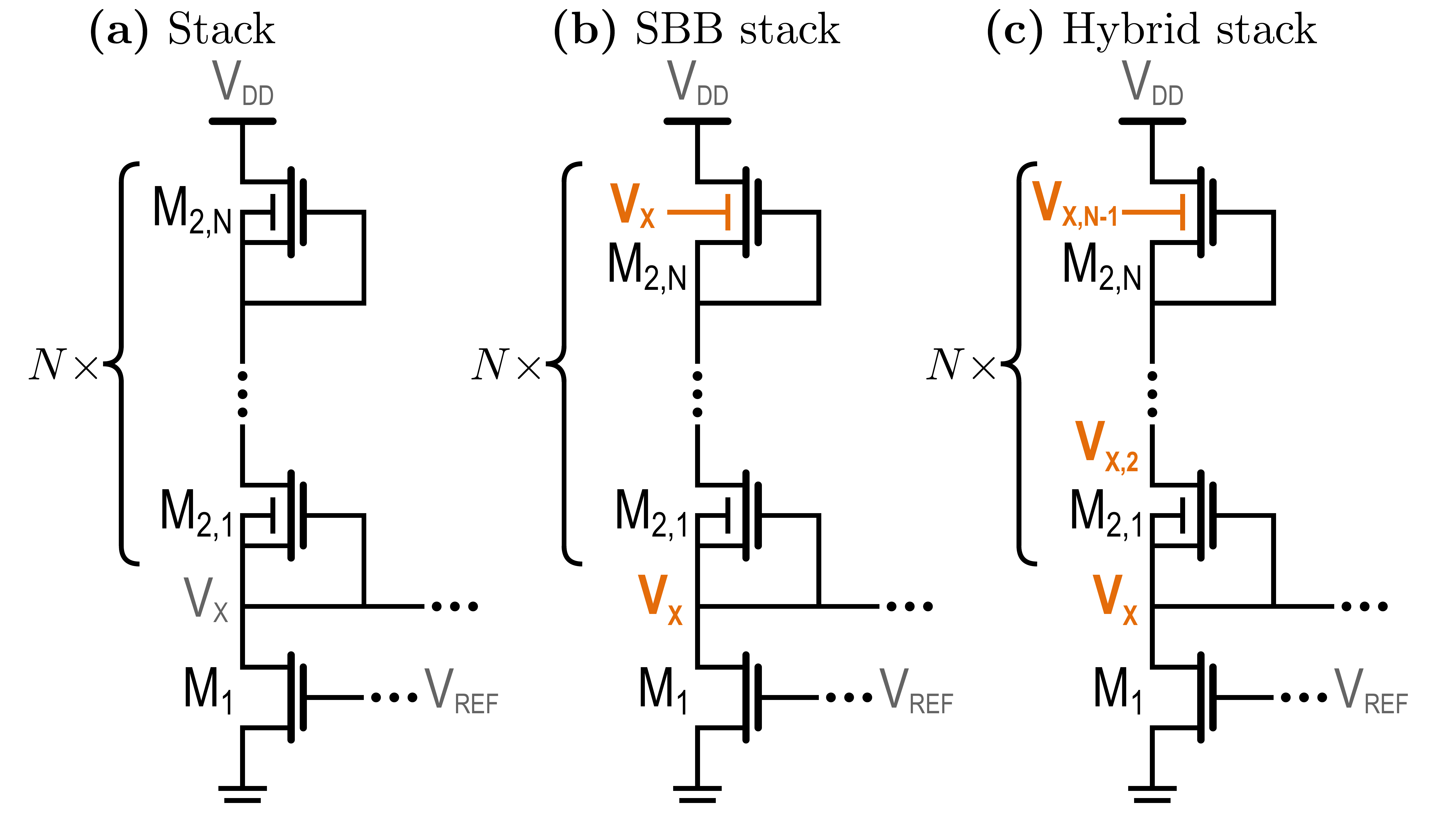}
	\caption{Line sensitivity enhancement techniques, illustrated here for nMOS topologies, using (a) stacking, (b) stacking with a shared body bias (SBB), and (c) hybrid stacking. The different body voltages in (b) and (c) are self-generated by the stack.}
	\label{fig:22_vref_design_LS_DC}
\end{figure}

\begin{figure}[!t]
	\centering
	\includegraphics[width=.431\textwidth]{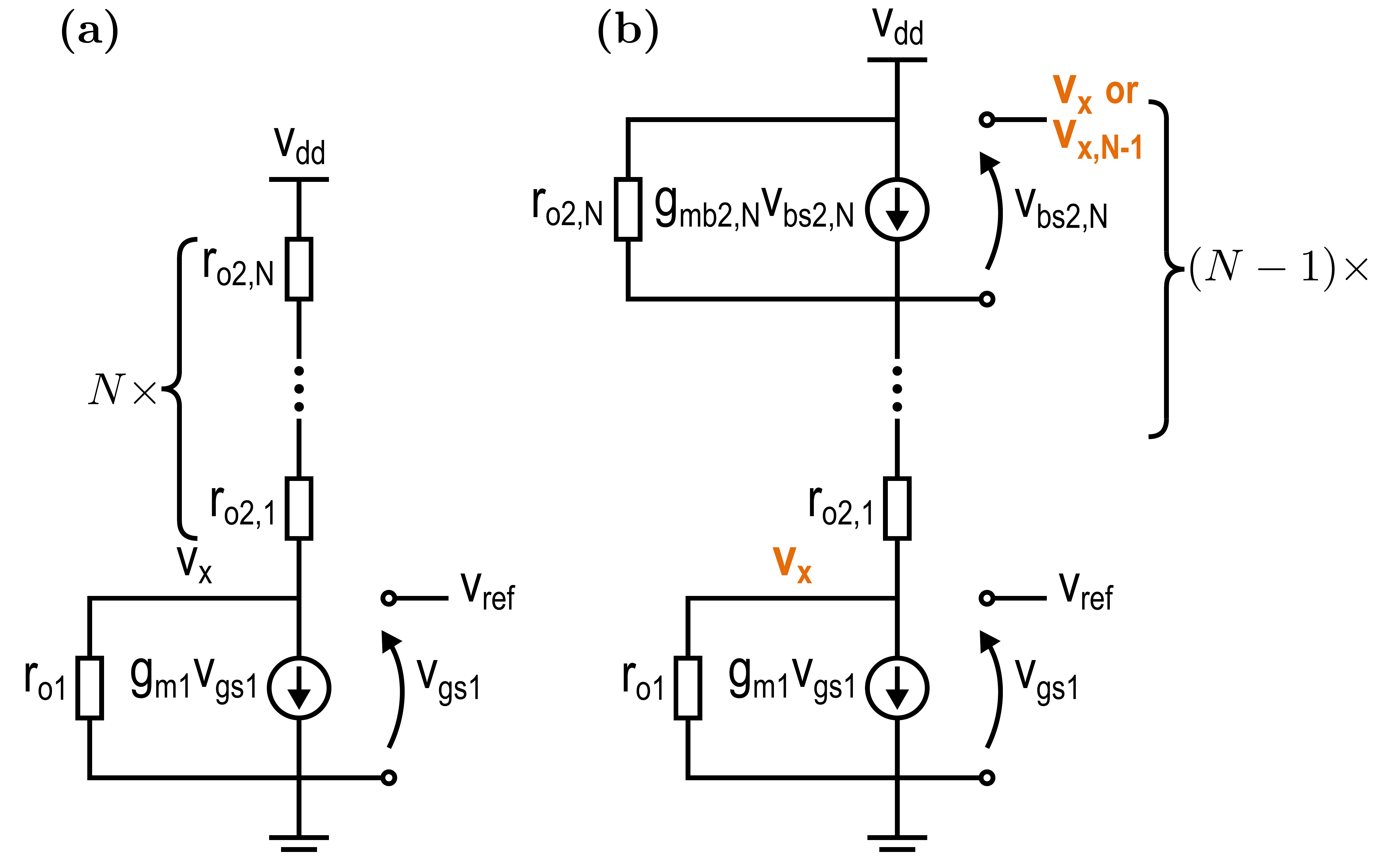}
	\caption{Small signal schematics of the LS enhancement circuits depicted in Fig.~\ref{fig:22_vref_design_LS_DC}, based on (a) stacking and (b) SBB/hybrid stacking.}
	\label{fig:23_vref_design_LS_small_signal}
\end{figure}

\begin{figure}[!t]
	\centering
	\includegraphics[width=.424\textwidth]{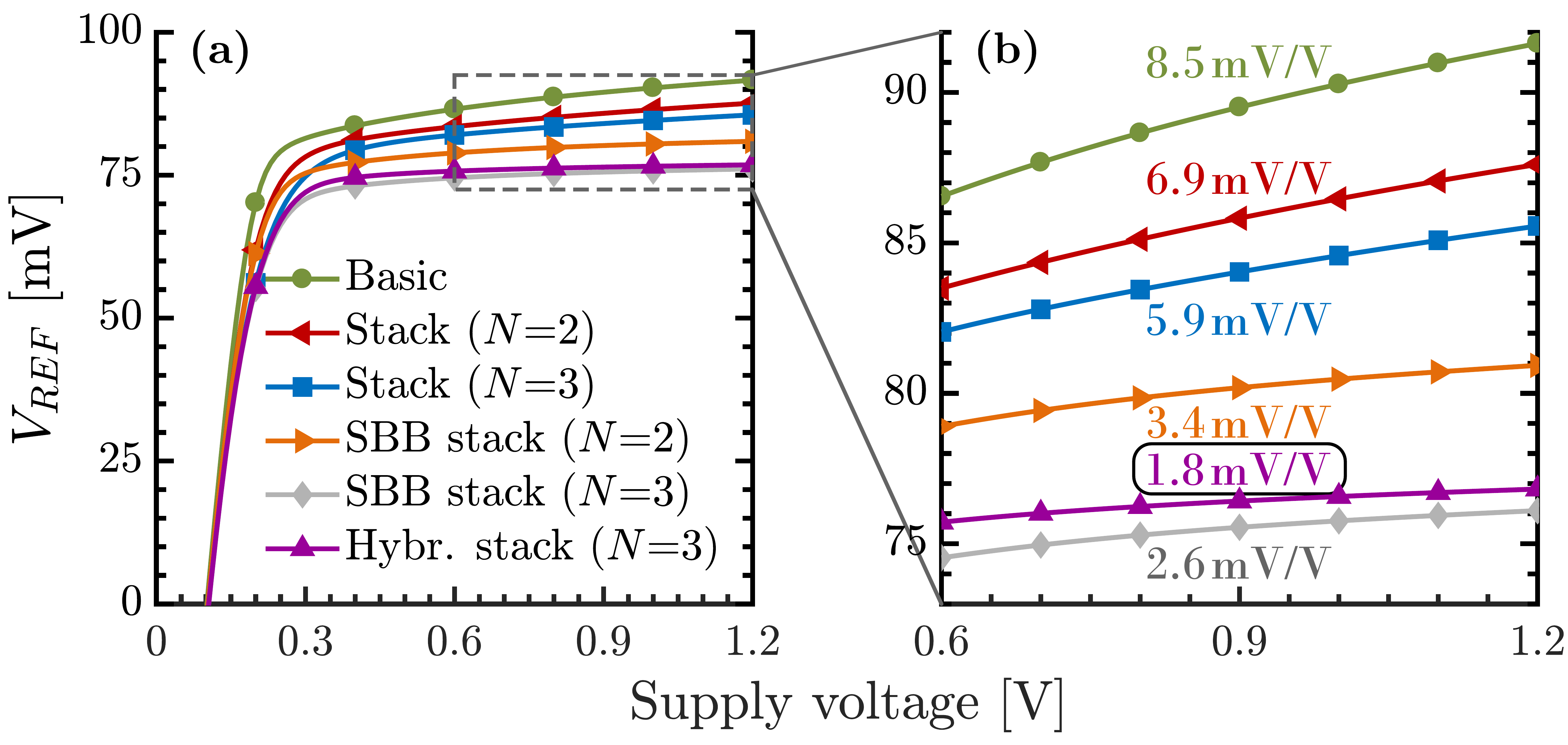}
	\caption{Comparison of the various LS enhancement techniques for a PTAT voltage reference implemented with \mbox{6-$\mu$m-length} LVT pMOS devices in \mbox{65-nm} bulk. The LS is measured from 0.6 to 1.2~V at 25$^\circ$C.}
	\label{fig:24_vref_design_stacking_results}
\end{figure}

As stated in (\ref{eq:vref_over_vdd}), the LS of the reference voltage is inversely proportional to the intrinsic gain, which is getting worse with each technology node as a result of increased output conductance \cite{Murmann_2006}. In addition, the problem posed by gate leakage and GIDL at low temperature does not allow to select a maximum-length transistor, which would have lead to a large $g_m/g_d$ and thus a lower LS. While this issue is not critical in \mbox{28-nm} FDSOI, as the intrinsic gain is larger than in bulk technologies \cite{Cathelin_2017}, it significantly degrades the LS in \mbox{65-nm} bulk. Therefore, several LS enhancement techniques are proposed in Fig.~\ref{fig:22_vref_design_LS_DC}: (a) stacking, (b) stacking with a shared body bias (SBB), and (c) hybrid stacking. These techniques are illustrated for nMOS topologies, albeit subsequent simulation results correspond to pMOS implementations, which limit the area overhead of using different body voltages by relying on nwells rather than on DTI or triple-well devices. Fig.~\ref{fig:24_vref_design_stacking_results} compares the various LS enhancement techniques to the basic solution, using a single transistor for $M_2$ and reaching an LS of 8.5~mV/V. First, employing a stack of $N$ devices with their body tied to their source [Figs.~\ref{fig:22_vref_design_LS_DC}(a) and \ref{fig:23_vref_design_LS_small_signal}(a)], the LS is expressed as
\begin{IEEEeqnarray}{L}
	\frac{v_{ref}}{v_{dd}} = \frac{\frac{g_{d2,1}}{N}}{g_{m1}+g_{d1}+\frac{g_{d2,1}}{N}} \simeq \frac{g_{d2,1}}{g_{m1}} \frac{1}{N},\IEEEnonumber%
\end{IEEEeqnarray}
under the assumptions that $g_{m1} \gg g_{d1},\:g_{d2,1}$ and $v_x = v_{ref}$. The LS decrease does not exactly scale as $1/N$, because the small signal parameters are impacted by the change of operation point due to stacking. Nevertheless, Fig.~\ref{fig:24_vref_design_stacking_results} shows an improved LS of 6.9 and 5.9~mV/V for a stacking of $N=2$ and 3 devices. Second, the LS using SBB stacking [Figs.~\ref{fig:22_vref_design_LS_DC}(b) and \ref{fig:23_vref_design_LS_small_signal}(b)], i.e., all bodies tied to $V_X$, or hybrid stacking [Figs.~\ref{fig:22_vref_design_LS_DC}(c) and \ref{fig:23_vref_design_LS_small_signal}(b)], i.e., the body of $M_{2,n}$ tied to the source of $M_{2,n-1}$, is expressed as
\begin{IEEEeqnarray}{CCL}
	\frac{v_{ref}}{v_{dd}} & \simeq & \frac{g_{d2,1}}{g_{m1}+g_{d1}} \frac{g_{d2,2}}{g_{mb2,2}+g_{d2,1}+g_{d2,2}},\IEEEnonumber\\
	\frac{v_{ref}}{v_{dd}} & \simeq & \frac{g_{d2,1}}{g_{m1}+g_{d1}} \frac{g_{d2,2}}{g_{mb2,2}+g_{d2,1}+g_{d2,2}} \frac{g_{d2,3}}{g_{mb2,3}+g_{d2,3}},\IEEEnonumber%
\end{IEEEeqnarray}
for stacks of $N=2$ and 3 devices, respectively. SBB and hybrid are equivalent for $N=2$ and lead to an LS of 3.4~mV/V, thanks to the second factor related to the body effect of $M_{2,2}$. Then, SBB and hybrid differ for $N=3$, reaching a 2.6- and \mbox{1.8-mV/V} LS, respectively. This difference comes from increased $g_{d2,1}$ and $g_{d2,2}$ in the SBB stack, arising from a $V_{DS}$ lower than 4$U_T$ for both $M_{2,1}$ and $M_{2,2}$. On the other hand, $M_{2,2}$ is saturated in the hybrid stack, lowering $g_{d2,2}$ and improving the LS. An hybrid stack with three devices is thus selected to implement the voltage reference.

\setlength{\tabcolsep}{2pt}
\begin{table*}[!t]
\centering
\caption{Comparison table of CWT (left) and PTAT (right, with TC highlighted in green) nA-range current references.}
\label{table:soa_nanoamp_range}
\resizebox{\linewidth}{!}{%
\begin{threeparttable}
\begin{scriptsize}
\begin{tabular}[t]{lcccccccccccccccc}
\toprule
& Cordova & Dong & Far & Hirose & Huang & Huang & Ji & Kayahan & Kim & Wang & Wang & Camacho & Camacho & \multicolumn{3}{c}{\textbf{Lefebvre}}\\
& \cite{Cordova_2017} & \cite{Dong_2017} & \cite{Far_2015} & \cite{Hirose_2010} & \cite{Huang_2010} & \cite{Huang_2020} & \cite{Ji_2017} & \cite{Kayahan_2013} & \cite{Kim_2016} & \cite{Wang_2019_VLSI} & \cite{Wang_2019_TCAS} & \cite{CamachoGaleano_2005} & \cite{CamachoGaleano_2008} & \multicolumn{3}{c}{\textbf{This work}}\\
\midrule
Publication & ISCAS & ESSCIRC & ROPEC & ESSCIRC & ISCAS & TCAS-II & ISSCC & TCAS-I & ISCAS & VLSI-DAT & TCAS-I & TCAS-II & ISCAS & \multicolumn{3}{c}{TCAS-I}\\
Year & 2017 & 2017 & 2015 & 2010 & 2010 & 2020 & 2017 & 2013 & 2016 & 2019 & 2019 & 2005 & 2008 & \multicolumn{3}{c}{2022}\\
\cmidrule(l){15-17}
Type of work & Sim. & Silicon & Sim. & Silicon & Sim. & Silicon & Silicon & Silicon & Sim. & Silicon & Silicon & Silicon & Silicon & Silicon & Sim.\tnote{$\blacktriangleleft$} & Sim.\tnote{$\blacktriangleleft$}\\
Number of samples & N/A & 32\tnote{$\dagger$} & N/A & 15 & N/A & 10 & 10 & 90 & N/A & 10 & 16 & 10 & 30 & 10 & N/A & N/A\\
\midrule
Technology & 0.18$\mu$m & 0.18$\mu$m & 0.18$\mu$m & 0.35$\mu$m & 0.18$\mu$m & 0.18$\mu$m & 0.18$\mu$m & 0.35$\mu$m & 0.13$\mu$m & 0.18$\mu$m & 0.18$\mu$m & 1.5$\mu$m & 0.18$\mu$m & 0.18$\mu$m & 65nm & 28nm\\
& & & & & & & & & & & & & & PDSOI & & FDSOI\\
$I_{REF}$ [nA] & 10.86 & 35.02 & 14 & 9.95 & 2.05 & 11.6 & 6.64 & 25 & 27 & 6.46 & 9.77 & 0.41 & 0.9 & 0.096 & 0.22 & 0.2\\
Power [nW] & 30.5 & 1.02 & 150 & 88.53 & 5.1 & 48.64 & 9.3 & 28500 & N/A & 15.8 & 28 & 2 & 2.34 & 0.282 & 0.80 & 0.68\\
& $@$0.9V & $@$1.5V & $@$1V & $@$1.3V & $@$0.85V & $@$0.8V & $@$N/A & $@$5V & & $@$0.85V & $@$0.7V & $@$1.1V & $@$0.65V & $@$0.55V & $@$0.7V & $@$0.74V\\
Area [mm$^2$] & 0.01 & 0.0169 & 0.0102 & \textcolor{ECS-Red}{\textbf{0.12}} & N/A & 0.054 & 0.055 & \textcolor{ECS-Blue}{\textbf{0.0053}} & N/A & 0.062 & 0.055 & 0.045 & 0.01 & \textcolor{ECS-Blue}{\textbf{0.0087}} & \textcolor{ECS-Blue}{\textbf{0.0054}} & \textcolor{ECS-Blue}{\textbf{0.0029}}\\
\midrule
Supply range [V] & 0.9 -- 1.8 & \textcolor{ECS-Red}{\textbf{1.5}} -- 2.5 & 1 -- 3.3 & \textcolor{ECS-Red}{\textbf{1.3}} -- 3 & 0.85 -- 2.2 & 0.8 -- 2 & \textcolor{ECS-Red}{\textbf{1.3}} -- 1.8 & N/A & 1.2 & 0.85 -- 2 & \textcolor{ECS-Blue}{\textbf{0.7}} -- 1.2 & 1.1 -- 3 & \textcolor{ECS-Blue}{\textbf{0.65}} -- 2 & \textcolor{ECS-Blue}{\textbf{0.55}} -- 1.8 & \textcolor{ECS-Blue}{\textbf{0.7}} -- 1.2 & \textcolor{ECS-Blue}{\textbf{0.74}} -- 1.8\\
LS [$\%$/V] & \textcolor{ECS-Blue}{\textbf{0.54}} & 3 & \textcolor{ECS-Blue}{\textbf{0.1}} & \textcolor{ECS-Blue}{\textbf{0.046}} & 1.35 & 1.08 & 1.16 & \textcolor{ECS-Red}{\textbf{150}} & N/A & \textcolor{ECS-Red}{\textbf{4.15}} & \textcolor{ECS-Blue}{\textbf{0.6}} & 3.5/\textcolor{ECS-Red}{\textbf{6}}\tnote{$\ast$} & \textcolor{ECS-Blue}{\textbf{0.2}} & 1.48 & \textcolor{ECS-Red}{\textbf{4.43}} & \textcolor{ECS-Blue}{\textbf{0.78}}\\
\midrule
Temperature range [$^\circ$C] & -20 -- 120 & -40 -- 120 & 0 -- 70 & -20 -- 80 & 0 -- 150 & -40 -- 120 & 0 -- 110 & 0 -- 80 & -30 -- 150 & -10 -- 100 & -40 -- 125 & -20 -- 70 & -70 -- 130 & -40 -- 85 & 0 -- 85 & -20 -- 85\\
TC [ppm/$^\circ$C] & 108 & 282 & \textcolor{ECS-Blue}{\textbf{20}} & 1200 & \textcolor{ECS-Blue}{\textbf{91}} & 169 \tnote{$\ddagger$} & 680/283\tnote{$\triangleleft$} & 128 (sim.) & 327 & 138\tnote{$\triangleright$} & 149.8 & 470/\textcolor{ECS-Green}{\textbf{2500}}\tnote{$\ast$} & \textcolor{ECS-Green}{\textbf{3000}} & \textcolor{ECS-Green}{\textbf{7500}} & \textcolor{ECS-Green}{\textbf{2930}} & \textcolor{ECS-Green}{\textbf{1095}}\\
\midrule
$I_{REF}$ var. (process) [$\%$] & 15.8 & 4.7 & N/A & N/A & 7.5 & +17.6/-10.3\tnote{$\diamond$} & N/A & 8/1.22\tnote{$\ast$} & 3.7 & N/A & +11.7/-8.7 & N/A & N/A & +14.5/-14.6 & +25.1/-2.0 & +5.3/6.3\\
$I_{REF}$ var. (mismatch) [$\%$] & \textcolor{ECS-Red}{\textbf{11.6}}\tnote{$\star$} & \textcolor{ECS-Blue}{\textbf{1.6}} & 5.8 & \textcolor{ECS-Red}{\textbf{14.1}} & N/A & 4.3 & 4.07/\textcolor{ECS-Blue}{\textbf{1.19}}\tnote{$\ast$} & \textcolor{ECS-Blue}{\textbf{1.4}} (sim.) & N/A & 3.33 & \textcolor{ECS-Blue}{\textbf{1.6}} & \textcolor{ECS-Red}{\textbf{10}} & 2.67 & \textcolor{ECS-Blue}{\textbf{1.66}} & 2.59 & 4.06\\
\midrule
Trimming type & 6b for & N/A & N/A & 3b for & N/A & 6b for & N/A & N/A & N/A & N/A & 5b for & N/A & N/A & N/A & N/A & N/A\\
& $V_S$ gen. & & & $I_N$/$I_P$ sub. & & TC/$V_{REF}$ & & & & & SCM width & & & & &\\
\midrule
Complexity & 25T & \textcolor{ECS-Blue}{\textbf{$>$4T+2C}} & 31T & \textcolor{ECS-Red}{\textbf{52T}} & \textcolor{ECS-Blue}{\textbf{10T}} & \textcolor{ECS-Red}{\textbf{23T+4C+2R}} & $>$18T+1R & 11T+1R & \textcolor{ECS-Blue}{\textbf{7T+1R}} & 12T & \textcolor{ECS-Red}{\textbf{23T+2C+2R}} & \textcolor{ECS-Blue}{\textbf{10T}} & 16T+1C & \textcolor{ECS-Blue}{\textbf{7T}} & \textcolor{ECS-Blue}{\textbf{9T}} & \textcolor{ECS-Blue}{\textbf{8T}}\\
Special components & \textcolor{ECS-Red}{\textbf{ZVT}} & No & No & No & No & Res. & Res., \textcolor{ECS-Red}{\textbf{1BJT}} & Res. & Res. & No & Res. & No & No & No & No & No\\
\bottomrule
\end{tabular}
\end{scriptsize}
\begin{footnotesize}
\begin{tablenotes}
	\item[$\ast$] Simulation and measurement results.
	\item[$\star$] Variability is due to the combined effects of process and mismatch variations. It is equal to 15.8~$\%$ for the untrimmed current reference and 11.6~$\%$ for the trimmed one.
	\item[$\dagger$] 16 dies for the TT process corner and 4 dies for each of the FF, SS, SF and FS process corners.
	\item[$\ddagger$] Average of the measurement results.
	\item[$\diamond$] Estimated from Fig.~3 of \cite{Huang_2020}.
	\item[$\triangleleft$] Untrimmed and trimmed results.
	\item[$\triangleright$] Minimum of the measurement results.
	\item[$\blacktriangleleft$] In this work, simulations refer to post-layout simulations including parasitic diodes.
\end{tablenotes}
\end{footnotesize}
\end{threeparttable}%
}
\end{table*}

\begin{table*}[!t]
\centering
\caption{Comparison table of $\mu$A-range temperature-independent current references.}
\label{table:soa_microamp_range}
\resizebox{\linewidth}{!}{%
\begin{threeparttable}
\begin{scriptsize}
\begin{tabular}[t]{lcccccccccccccccc}
\toprule
& Azcona & Bendali & Crupi & Eslampanah & Fiori & Lee & Liu & Osipov & Serrano & Wang & Wu & Yang & Yoo & \multicolumn{3}{c}{\textbf{Lefebvre}}\\
& \cite{Azcona_2014} & \cite{Bendali_2007} & \cite{Crupi_2018} & \cite{Eslampanah_2017} & \cite{Fiori_2005} & \cite{Lee_2012} & \cite{Liu_2010} & \cite{Osipov_2019} & \cite{Serrano_2008} & \cite{Wang_2017} & \cite{Wu_2015} & \cite{Yang_2009} & \cite{Yoo_2007} & \multicolumn{3}{c}{\textbf{This work}}\\
\midrule
Publication & ISCAS & TCAS-I & & ISCAS & TCAS-II & JSSC & ISCAS & TCAS-I & JSSC & TVLSI & CICC & ASSCC & & \multicolumn{3}{c}{TCAS-I}\\
Year & 2014 & 2007 & 2017 & 2017 & 2005 & 2012 & 2010 & 2019 & 2008 & 2017 & 2015 & 2009 & 2007 & \multicolumn{3}{c}{2022}\\
\cmidrule(l){15-17}
Type of work & Sim. & Silicon & Silicon & Silicon & Sim. & Silicon & Sim. & Silicon & Silicon & Silicon & Silicon & Silicon & Silicon & Silicon & Sim.\tnote{$\triangleright$} & Sim.\tnote{$\triangleright$}\\
Number of samples & N/A & 19 & 45 & 5 & N/A & 10 & N/A & 10 & N/A & 12 & N/A & N/A & N/A & 10 & N/A & N/A\\
\midrule
Technology & 0.18$\mu$m & 0.18$\mu$m & 0.18$\mu$m & 0.18$\mu$m & 0.35$\mu$m & 0.18$\mu$m & 0.18$\mu$m & 0.35$\mu$m & 0.5$\mu$m & 65nm & 0.18$\mu$m & 0.35$\mu$m & 0.25$\mu$m & 0.18$\mu$m & 65nm & 28nm\\
& & & & & BiCMOS & & & & & & & & & PDSOI & & FDSOI\\
$I_{REF}$ [$\mu$A] & 0.5 & 144.3 & 0.34 & 1.5 & 13.65 & 7.81 & 10 & 1.05 & 16 -- 50 & 9.42 & 20 & 16.5 & 10.45 & 0.99 / 1.09 & 1 & 1\\
Power [$\mu$W] & 1.98 & 83 & 0.21 & 1.38 & 68.25 & 1.4 & 80 & 3.8 & N/A & 7.18 & 48 & N/A & 77 & 0.64 / 0.71 & 0.68 & 0.80\\
& $@$1.2V & $@$1V & $@$0.45V & $@$0.8V & $@$2.5V & $@$1V & $@$2V & $@$2V & & $@$1.4V & $@$2.4V & & $@$1.1V & $@$0.65V & $@$0.7V & $@$0.8V\\
Area [mm$^2$] & N/A & \textcolor{ECS-Red}{\textbf{0.213}}\tnote{$\star$} & \textcolor{ECS-Blue}{\textbf{0.00075}} & 0.003 & 0.0042 & 0.023 & N/A & 0.057 & 0.06\tnote{$\triangleleft$} & \textcolor{ECS-Red}{\textbf{0.089}} & 0.005 & 0.0576 & \textcolor{ECS-Blue}{\textbf{0.002}} & 0.0034 / 0.0043 & \textcolor{ECS-Blue}{\textbf{0.002}} & \textcolor{ECS-Blue}{\textbf{0.00044}}\\
\midrule
Supply range [V] & 1.2 -- 3 & 1 & \textcolor{ECS-Blue}{\textbf{0.45}} -- 1.8 & \textcolor{ECS-Blue}{\textbf{0.8}} -- 2 & \textcolor{ECS-Red}{\textbf{2.5}} & 1 -- 1.2 & 2 -- 3 & 2 -- 3.6 & \textcolor{ECS-Red}{\textbf{2.3}} -- 3.3 & 1.3 -- 2.5 & \textcolor{ECS-Red}{\textbf{2.4}} -- 3 & 2 -- 3.3 & 1.1 -- 3 & \textcolor{ECS-Blue}{\textbf{0.65}} -- 1.8 & \textcolor{ECS-Blue}{\textbf{0.7}} -- 1.2 & \textcolor{ECS-Blue}{\textbf{0.8}} -- 1.8\\
LS [$\%$/V] & 0.69 & N/A & \textcolor{ECS-Red}{\textbf{3.9}} & N/A & \textcolor{ECS-Blue}{\textbf{0.4}} & N/A\tnote{$\diamond$} & 3 & 2.73 & $<$1 & \textcolor{ECS-Blue}{\textbf{0.018}} & 0.5 & N/A & \textcolor{ECS-Blue}{\textbf{0.17}} & \textcolor{ECS-Blue}{\textbf{0.20 / 0.21}} & \textcolor{ECS-Red}{\textbf{4.38}} & 0.57\\
\midrule
Temperature range [$^\circ$C] & -40 -- 120 & 0 -- 100 & 0 -- 80 & -40 -- 110 & -30 -- 100 & 0 -- 100 & -20 -- 120 & -45 -- 125 & 0 -- 80 & -30 -- 90 & -40 -- 80 & -20 -- 100 & 0 -- 120 & -40 -- 85 & 0 -- 125 & -40 -- 125\\
TC [ppm/$^\circ$C] & 119 & 185 & \textcolor{ECS-Red}{\textbf{578}} & \textcolor{ECS-Red}{\textbf{571}} & \textcolor{ECS-Blue}{\textbf{28}} & \textcolor{ECS-Blue}{\textbf{24.9}} & 170 & 143 (min.) & $<$130 & \textcolor{ECS-Blue}{\textbf{86 (min.)}} & 130 & 280\tnote{$\ast$} & \textcolor{ECS-Red}{\textbf{720}} & 290 / \textcolor{ECS-Blue}{\textbf{38}}\tnote{$\blacktriangleleft$} & \textcolor{ECS-Blue}{\textbf{37.33}}\tnote{$\blacktriangleright$} & \textcolor{ECS-Blue}{\textbf{38.42}}\tnote{$\blacktriangleright$}\\
& & & & & & & & 247 (avg.) & & 119 (avg.) & & & & & &\\
\midrule
$I_{REF}$ var. (process) [$\%$] & 1.3\tnote{$\ast$} & 7\tnote{$\dagger$} & N/A & 8.5 & N/A & 4.5 & 2.14\tnote{$\dagger$} & N/A & N/A & 5.13 & N/A & 0.52\tnote{$\ast$} & N/A & +33.3/-25.9 & +46.7/-22.6 & +9.8/-10.1\\
$I_{REF}$ var. (mismatch) [$\%$] & N/A & & 2.7\tnote{$\ddagger$} & N/A & N/A & N/A & & 3.9 & \textcolor{ECS-Blue}{\textbf{$<$0.02}} & \textcolor{ECS-Blue}{\textbf{0.4}} & \textcolor{ECS-Red}{\textbf{14}} & N/A & N/A & \textcolor{ECS-Blue}{\textbf{0.65 / 0.87}} & \textcolor{ECS-Blue}{\textbf{0.68}} & \textcolor{ECS-Blue}{\textbf{0.85}}\\
\midrule
Trimming type & 2b for TC & N/A & 4b for TC & N/A & N/A & N/A & N/A & N/A & Floating charge & N/A & N/A & 10b for TC & N/A & 4b for TC & N/A & N/A\\
& 2b for $I_{REF}$ & & 4b for $I_{REF}$ & & & & & & for TC, $I_{REF}$ & & & 8b for $I_{REF}$ & & & &\\
\midrule
Complexity & \textcolor{ECS-Red}{\textbf{39T+2R}} & \textcolor{ECS-Red}{\textbf{30T+2C+5R}} & \textcolor{ECS-Blue}{\textbf{2T}} & \textcolor{ECS-Blue}{\textbf{4T+2R}} & \textcolor{ECS-Blue}{\textbf{5T+2R}} & 12T+1C+2R & 38T & \textcolor{ECS-Blue}{\textbf{5T+1R}} & \textcolor{ECS-Red}{\textbf{$>$26T+5C+4R}} & $>$28T+3R & 15T+2R & 27T+2R & 18T+3R & \textcolor{ECS-Blue}{\textbf{4/12T+1R}} & \textcolor{ECS-Blue}{\textbf{5T+1R}} & \textcolor{ECS-Blue}{\textbf{4T+1R}}\\
Special components & Res. & Res., \textcolor{ECS-Red}{\textbf{2BJT}} & No & Res. & Res. & Res. & \textcolor{ECS-Red}{\textbf{3BJT}} & Res. & Res. & Res. & Res., \textcolor{ECS-Red}{\textbf{2BJT}} & Res., \textcolor{ECS-Red}{\textbf{5BJT}} & Res. & Res. & Res. & Res.\\
\bottomrule
\end{tabular}
\end{scriptsize}
\begin{footnotesize}
\begin{tablenotes}
	\item[$\ast$] After trimming.
	\item[$\star$] Estimated from the microphotograph in Fig.~6 of \cite{Bendali_2007}.
	\item[$\dagger$] Variability is due to the combined effects of process and mismatch variations.
	\item[$\ddagger$] Before trimming.
	\item[$\diamond$] Line sensitivity is not measured because the current reference is supplied by a bandgap reference (BGR).
	\item[$\triangleleft$] The total area is split between the current reference (0.015~mm$^2$) and the charge pumps (0.045~mm$^2$).
	\item[$\triangleright$] In this work, simulations refer to post-layout simulations including parasitic diodes.
	\item[$\blacktriangleleft$] Average of the measurement results over 10 samples.
	\item[$\blacktriangleright$] Median of 10$^4$ local mismatch Monte-Carlo runs.
\end{tablenotes}
\end{footnotesize}
\end{threeparttable}%
}
\end{table*}

\section{Comparison to the State of the Art}
\label{sec:comparison_to_the_state_of_the_art}
\begin{figure}[!t]
	\centering
	\includegraphics[width=.424\textwidth]{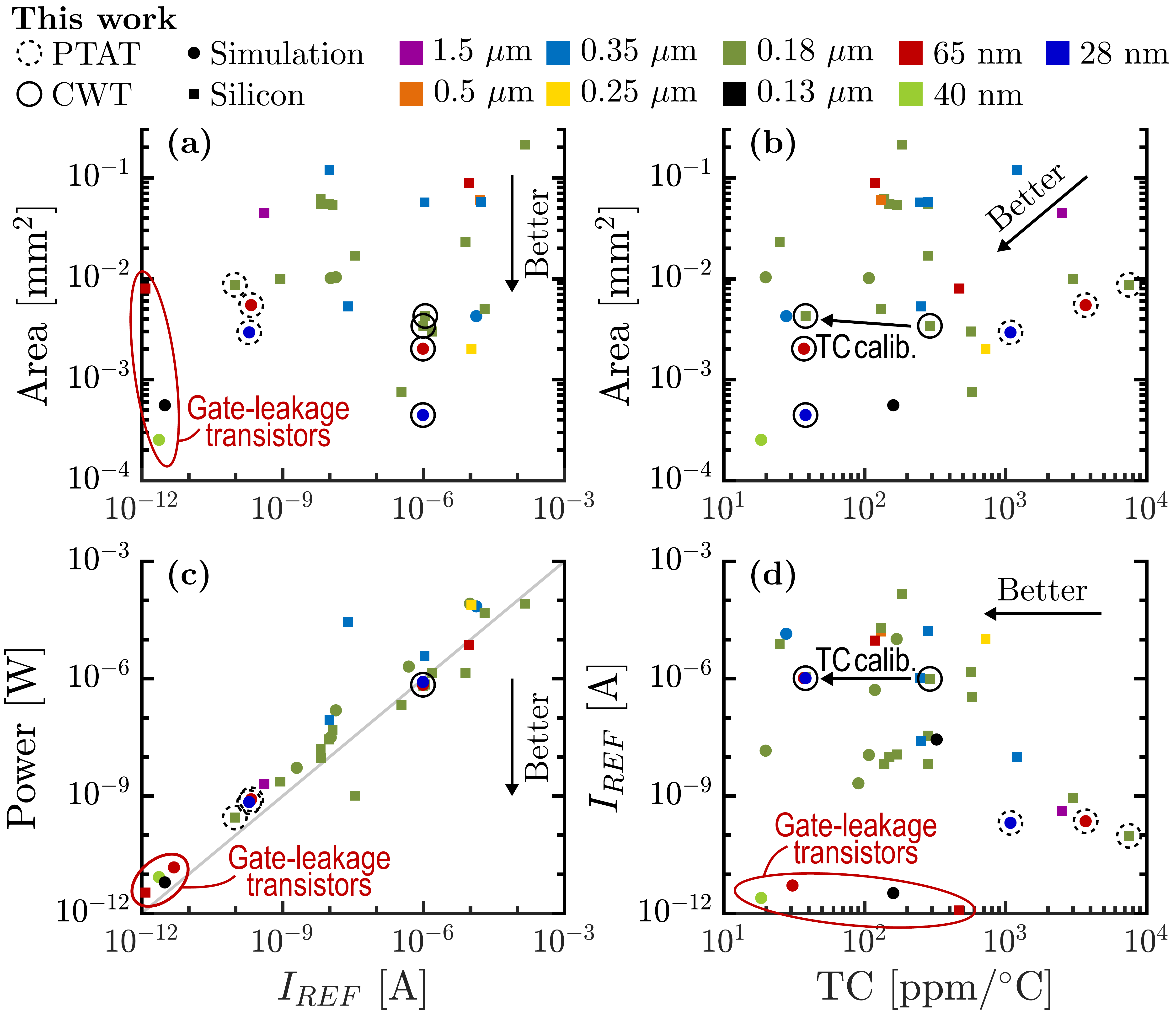}
	\caption{Illustration of the trade-offs between (a) area and $I_{REF}$, (b) area and TC, (c) power and $I_{REF}$, and (d) $I_{REF}$ and TC, based on the state of the art of nA- and $\mu$A-range current references and a few additional references.}
	\label{fig:25_comparison_to_soa}
\end{figure}
This section compares our work to the state of the art of nA-range current references in Table~\ref{table:soa_nanoamp_range}, and $\mu$A-range temperature-independent ones in Table~\ref{table:soa_microamp_range}. These tables are complemented by Fig.~\ref{fig:25_comparison_to_soa}, representing some of the most important trade-offs for current references. For both PTAT and CWT designs, measurements are reported for the \mbox{0.18-$\mu$m} PDSOI references. In addition, we present post-layout simulations in \mbox{0.18-$\mu$m} PDSOI, \mbox{65-nm} bulk and \mbox{28-nm} FDSOI, while the vast majority of other works limit themselves to older technology nodes, typically \mbox{0.18-$\mu$m} or above.\\
\indent First, our PTAT design reaches the lowest current and power among nA-range references (Table~\ref{table:soa_nanoamp_range}), as depicted in Fig.~\ref{fig:25_comparison_to_soa}(c), and is only outperformed by references using gate-leakage transistors. For the \mbox{0.18-$\mu$m} reference, this corresponds to a 4.3$\times$ current and 7.1$\times$ power reduction compared to \cite{CamachoGaleano_2005}, respectively, enabled by a cut of $V_{DD,min}$ from 1.1 to 0.55~V. $I_{REF}$ is slightly larger for the 65- and \mbox{28-nm} references (0.2\hspace{2pt}/\hspace{2pt}0.22~nA), with a power consumption of 0.8\hspace{2pt}/\hspace{2pt}0.88~nW due to the larger $I_{REF}$ and $V_{DD,min}$. Moreover, our \mbox{0.18-$\mu$m} reference has the lowest $V_{DD,min}$ (0.55~V) among references in the nA range, followed by \cite{CamachoGaleano_2008}, \cite{Wang_2019_TCAS} and our 65- and \mbox{28-nm} designs in the \mbox{0.65-to-0.75-V} range. Regarding LS, our \mbox{0.18-$\mu$m} and \mbox{28-nm} designs reach competitive values of 1.48~$\%$/V and 0.78~$\%$/V allowed by a large intrinsic gain in these technologies, while the \mbox{65-nm} one exhibits a \mbox{4.43-$\%$/V} LS. Nonetheless, LS is significantly reduced from above 15~$\%$/V down to 4.43~$\%$/V as a result of using an hybrid stack of three devices (Section~\ref{subsec:implementation_in_scaled_technologies_B}). Then, the proposed topology is strikingly simple compared to other works, with only seven to nine transistors: two to four for the voltage reference and five for the SCM. Furthermore, it can be implemented with a single transistor type, and does not require any BJTs, startup circuit or trimming, thus resulting in a low silicon area ranging from 8700~$\mu$m$^2$ in \mbox{0.18-$\mu$m} down to 2900~$\mu$m$^2$ in \mbox{28-nm} [Fig.~\ref{fig:25_comparison_to_soa}(a)]. Only \cite{Dong_2017} and \cite{Kim_2016} use simpler structures, but either occupy a 1.9$\times$ larger or unreported area. Speaking of area, \cite{Kayahan_2013} is the closest competitor, but it suffers from a prohibitive power of 28.5~$\mu$W for a reference current of 25~nA. As far as temperature range is concerned, the lower limit in 65- and \mbox{28-nm} designs is due to gate leakage and GIDL, respectively, while the upper limit of 85$^\circ$C shared by all designs originates from the leakage in parasitic nwell/psub diodes at high temperature. The temperature dependence of all three proposed references follows the specific sheet current, meaning that TC is not a relevant comparison criterion, but also that the main drawback of the proposed topology is that it suffers from process variations, although additional $I_{REF}$ calibration is possible. Regarding the variability due to local mismatch, it is limited to 1.66~$\%$ in the \mbox{0.18-$\mu$m} reference, which is among the lowest values reported in Table~\ref{table:soa_nanoamp_range}, while 65- and \mbox{28-nm} references exhibit a larger variability due to a larger standard deviation of $V_{REF}$.\\
\indent Second, our CWT current references (Table~\ref{table:soa_microamp_range}) also present a simple structure, consisting of a single resistor and four to five transistors, depending on the voltage reference implementation. Therefore, the silicon area is modest compared to other $\mu$A-range references [Fig.~\ref{fig:25_comparison_to_soa}(a)], with 3410\hspace{2pt}/\hspace{2pt}4270~$\mu$m$^2$ in \mbox{0.18-$\mu$m}, down to a best-in-class \mbox{440-$\mu$m$^2$} area in \mbox{28-nm}. \cite{Crupi_2018} has both a small \mbox{750-$\mu$m$^2$} area and a simple structure, as it relies on a 2T voltage reference to bias a transistor gate [Fig.~\ref{fig:1_previous_works}(b)], but it suffers from degraded LS and TC, and has a limited \mbox{0-to-80$^\circ$C} temperature range. \cite{Eslampanah_2017}, \cite{Fiori_2005} and \cite{Osipov_2019} rely on a handful of components, but \cite{Eslampanah_2017} has a large \mbox{0.003-mm$^2$} area, and \cite{Fiori_2005, Osipov_2019} require a substantial supply voltage above 2~V to operate. \cite{Yoo_2007} occupies a limited \mbox{2000-$\mu$m$^2$} area, but its \mbox{720-ppm/$^\circ$C} TC is relatively poor, and the 18T+3R structure will likely lead to prohibitive variability due to local mismatch. Regarding TC, our three references feature a very low \mbox{38-ppm/$^\circ$C} TC for this current level [Fig.~\ref{fig:25_comparison_to_soa}(d)], and offer one of the best trade-offs between TC and silicon area [Fig.~\ref{fig:25_comparison_to_soa}(b)]. Other works in Table~\ref{table:soa_microamp_range} present a low TC, such as \cite{Fiori_2005} with 28 ppm/$^\circ$C, which requires a \mbox{2.5-V} supply voltage and consumes a considerable \mbox{68.3-$\mu$W} power, and \cite{Lee_2012} with 25~ppm/$^\circ$C, which necessitates a bandgap reference voltage to have a decent LS and occupies 5.3$\times$ more area than our largest design. As for other metrics, LS amounts to 0.2~$\%$/V and 0.57~$\%$/V in \mbox{0.18-$\mu$m} and \mbox{28-nm}, but increases up to 4.38~$\%$/V in \mbox{65-nm}, despite the use of an SBB stack of two devices. On one side, the lower temperature limit is -40$^\circ$C, except in \mbox{65-nm} where it rises to 0$^\circ$C due to gate leakage. On the other side, the upper limit is 125$^\circ$C, except in \mbox{0.18-$\mu$m} where it is limited to 85$^\circ$C by the measurement equipment. Similarly to their PTAT counterparts, our CWT references are significantly affected by process variations, an issue that can be alleviated through a calibration of $I_{REF}$. Nevertheless, variability due to local mismatch is comprised between 0.65 and 0.87~$\%$, which is among the best in the state of the art. Finally, our CWT references avoid the disadvantages of gate-leakage ones, which generate a pA-range current with limited power and area but can only be implemented in advanced technology nodes, suffer from a limited temperature range, and do not retain their area and power advantages for larger current levels.

\section{Conclusion}
\label{sec:conclusion}
In this work, we proposed two novel current reference topologies sharing two key ideas: (i) the generation of a voltage reference by a 2T ULP structure and (ii) its buffering onto a V-to-I converter by a single transistor. These references are fabricated in a \mbox{0.18-$\mu$m} PDSOI process. First, a nA-range PTAT current is obtained by biasing an SCM with a PTAT voltage. It generates a \mbox{0.096-nA} current, the lowest to date for current references without gate-leakage transistors, while consuming only 0.28~nW at 0.55~V and 25$^\circ$C. Second, a $\mu$A-range CWT current is obtained by biasing a polysilicon resistor with a matched-TC voltage. It generates either a \mbox{0.99-$\mu$A} current with a \mbox{290-ppm/$^\circ$C} CTAT TC, or a \mbox{1.09-$\mu$A} current with a \mbox{38-ppm/$^\circ$C} TC, using a 4-bit calibration of the 2T voltage reference width ratio. Both references exhibit a decent LS (1.48\hspace{2pt}/\hspace{2pt}0.21~$\%$/V), a low $V_{DD,min}$ (0.55\hspace{2pt}/\hspace{2pt}0.65~V), a low variability due to local mismatch (1.66\hspace{2pt}/\hspace{2pt}0.87~$\%$), while relying on simple 7T\hspace{2pt}/\hspace{2pt}4T+1R structures without any startup circuit and occupying a limited silicon area of 8700\hspace{2pt}/\hspace{2pt}4270~$\mu$m$^2$. The main drawback of such architectures is their sensitivity to process variations. Furthermore, we discuss the challenges posed by gate leakage, GIDL, and declining intrinsic gain to the implementation of 2T voltages references in scaled 65- and \mbox{28-nm} technologies. We demonstrate that a proper selection of the transistor type and sizes, together with the use of LS enhancement techniques, can mitigate these non-idealities.


%

\appendices

\section*{Acknowledgment}
We thank Pierre Gérard for the measurement testbench, Eléonore Masarweh for the microphotograph, Denis Flandre for fruitful discussions, and colleagues for their proofreading.
\ifCLASSOPTIONcaptionsoff
  \newpage
\fi



%
\bibliographystyle{IEEEtran}
\bibliography{Lefebvre_TCASI_2021}

%

\begin{IEEEbiography}[{\includegraphics[width=1in,height=1.25in,clip,keepaspectratio]{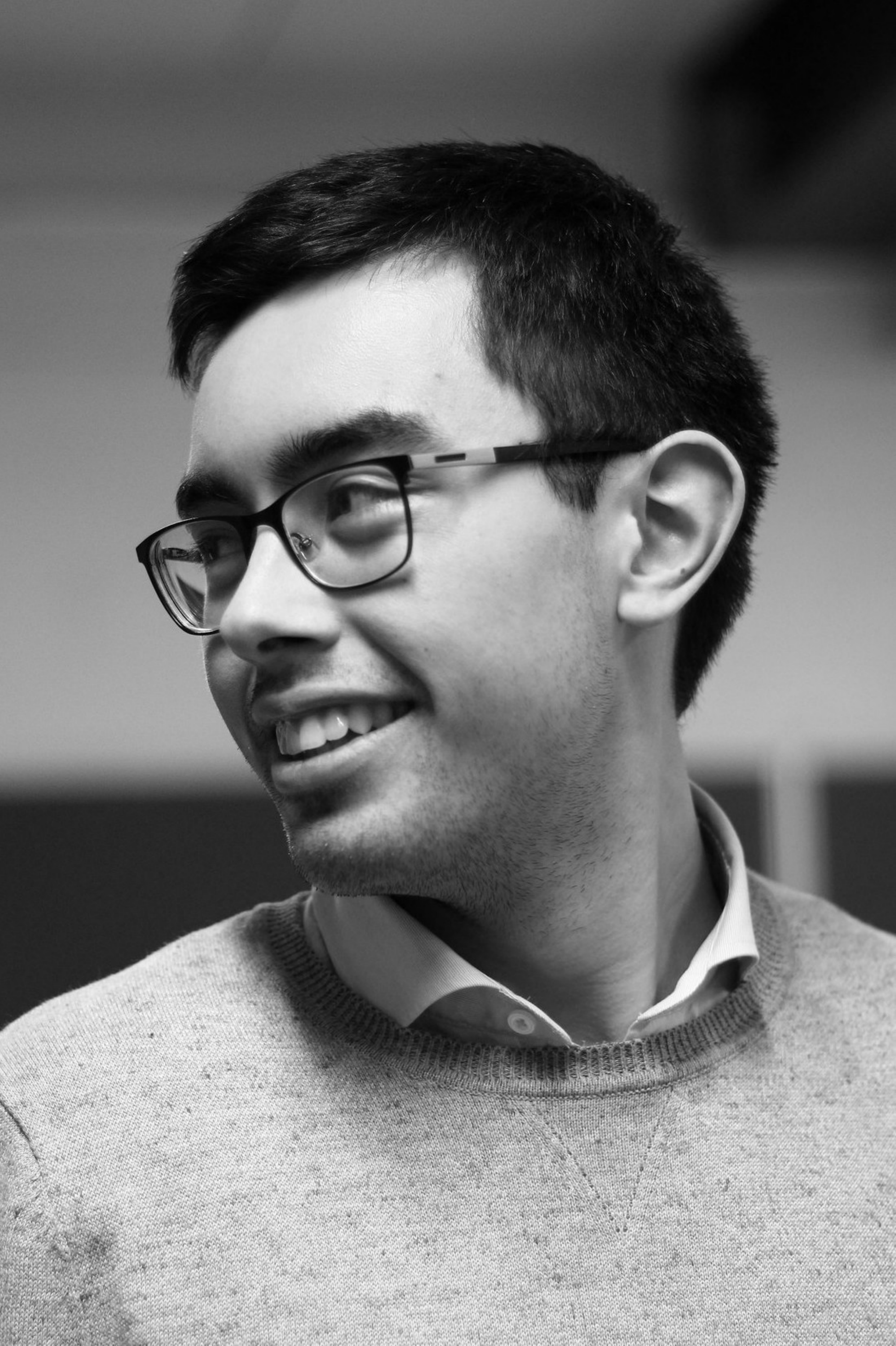}}]{Martin Lefebvre} (S’17) received the M.Sc. degree (summa cum laude) in electromechanical engineering from the Universit\'e catholique de Louvain (UCLouvain), Louvain-la-Neuve, Belgium, in 2017, where he is currently pursuing the Ph.D. degree, under the supervision of Prof. D. Bol. His current research interests include hardware-aware machine learning algorithms, low-power mixed-signal vision chips for embedded image processing, and ultra-low-power current reference architectures. He serves as a reviewer for various conferences and journals, including  IEEE Trans. on Biomed. Circuits and Syst., IEEE Trans. on VLSI Syst., Int. Symp. on Circuits and Syst. and Asia Pacific Conf. on Circuits and Syst.
\end{IEEEbiography}
\vspace{-1cm}

\begin{IEEEbiography}[{\includegraphics[width=1in,height=1.25in,clip,keepaspectratio]{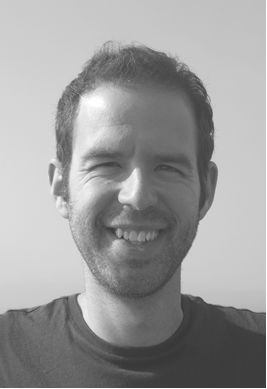}}]{David Bol}
received the Ph.D degree in Engineering Science from the Universit\'e catholique de Louvain (UCLouvain), Louvain-la-Neuve, Belgium in 2008. In 2005, he was a visiting Ph.D student at the CNM National Centre for Microelectronics, Sevilla, Spain, in advanced logic design. In 2009, he was a postdoctoral researcher at intoPIX, Louvain-la-Neuve, Belgium, in low-power image processing. In 2010, he was a visiting postdoctoral researcher at the UC Berkeley Laboratory for Manufacturing and Sustainability, Berkeley, CA, in life-cycle assessment of the semiconductor environmental impacts. He is now an Associate Professor at UCLouvain. In 2015, he participated to the creation of e-peas semiconductors, Louvain-la-Neuve, Belgium.
Prof. Bol leads the Electronic Circuits and Systems (ECS) group focused on ultra-low-power design of integrated circuits for the IoT and biomedical applications including computing, power management, sensing and wireless communications. He is engaged in a social-ecological transition in the field of ICT research. He co-teaches four M.S. courses on digital, analog and mixed-signal ICs, sensors and systems, with two B.S. courses including the course on Sustainable Development and Transition.
Prof. Bol has authored or co-authored more than 120 technical papers and conference contributions and holds three delivered patents. He (co-)received four Best Paper/Poster/Design Awards in IEEE conferences (ICCD 2008, SOI Conf. 2008, FTFC 2014, ISCAS 2020). He serves as a reviewer for various journals and conferences such as IEEE J. of Solid-State Circuits, IEEE Trans. on VLSI Syst., IEEE Trans. on Circuits and Syst. I/II. Since 2008, he presented several invited papers and keynote tutorials in international conferences including a forum presentation at IEEE ISSCC 2018.
On the private side, he pioneered the parental leave for male professors in his faculty to spend time connecting to nature with his family. 
\end{IEEEbiography}





\end{document}